\documentclass[manuscript,nonacm]{acmart}

\ifdefined\pdfcompresslevel
\fi

\usepackage{soul}
\usepackage{graphicx}

\usepackage{amssymb,mathtools}
\usepackage{multirow}
\usepackage{array}
\usepackage{subcaption}

\usepackage[ruled,vlined,linesnumbered]{algorithm2e}
\SetKwInput{KwIn}{Input}\SetKwInput{KwOut}{Output}
\SetKwFor{ForEach}{for each}{do}{end}\DontPrintSemicolon

\usepackage{booktabs,tabularx,seqsplit,makecell}
\newcolumntype{Y}{>{\raggedright\arraybackslash}X}
\usepackage{enumitem}

\usepackage{xspace}
\usepackage{xcolor}
\usepackage{wrapfig}

\let\originalcite\cite
\renewcommand{\cite}[1]{%
  \StrSubstitute{#1}{,chatmuse}{,zhou2026chatmuse}[\mappedcitekeys]%
  \originalcite{\mappedcitekeys}%
}

\newcommand*{\eg}{e.g.\@\xspace}
\newcommand*{\ie}{i.e.\@\xspace}

\newcommand{\ssection}[1]{\noindent{\textbf{#1}}}

\definecolor{fcolor}{rgb}{.729, .784, .827}

\newcommand{\vocaleyes}{\textsc{VocalEyes}}
\colorlet{participant}{black!70}

\begin{document}

\title{VocalEyes: Speaker-Aware Augmented Reality Captioning through In-Conversation Registration}

\author{Yuxiao Wang}
\affiliation{%
  \institution{University of Texas at Dallas}
  \city{Richardson}
  \state{TX}
  \country{USA}
}
\email{yuxiao.wang@utdallas.edu}

\author{Xulong Tang}
\affiliation{%
  \institution{University of Texas at Dallas}
  \city{Richardson}
  \state{TX}
  \country{USA}
}
\email{xulong.tang@utdallas.edu}

\author{Chen Chen}
\affiliation{%
  \institution{Florida International University}
  \city{Miami}
  \state{FL}
  \country{USA}
}
\email{che.chen@fiu.edu}

\author{Rawan Alghofaili}
\affiliation{%
  \institution{University of Texas at Dallas}
  \city{Richardson}
  \state{TX}
  \country{USA}
}
\email{rawan@utdallas.edu}

\begin{abstract}
Co-located augmented reality~(AR) captions make speech readable, but they can separate an utterance from the person who produced it. In unfamiliar groups, losing that source complicates immediate responses and later review: users must recover not only what was said, but also who said it. Conventional diarization returns anonymous clusters, while speaker recognition typically assumes pre-meeting enrollment. We built \vocaleyes{}, a speaker-aware AR captioning system that creates named voice profiles from natural self-introductions. The interface coordinates speaker-attributed captions, a fixed profile card, and a visual cue that marks the articulating face. In a controlled within-subjects study with $20$ participants who reported typical hearing, \vocaleyes{} identified speakers with $88.0\%$ accuracy and increased participant speaker-tracking accuracy from $47.2\%$ with caption-only AR to $87.3\%$ with the complete interface. Participants also reported lower workload with the complete interface. These findings show how in-conversation registration can preserve speaker attribution across live captions and meeting records in scripted small-group meetings.
\end{abstract}

\begin{teaserfigure}
  \centering
  \includegraphics[width=0.75\textwidth]{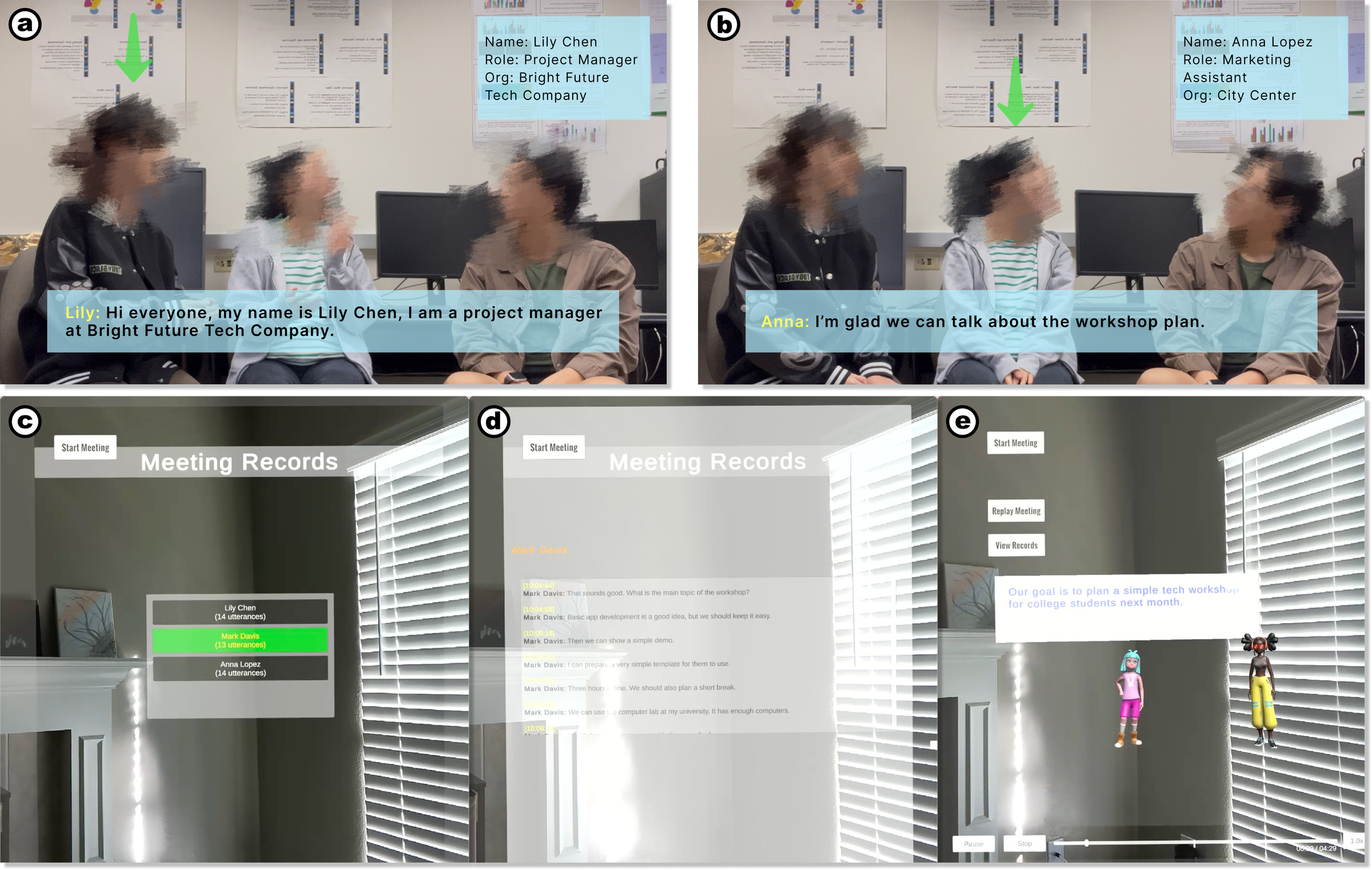}
  \vspace{-4mm}
  \caption{The \vocaleyes{} identity lifecycle. A natural self-introduction registers a named voice profile; later turns retain that name in the caption while a separate marker indicates the face detected as articulating (a--b). The same attribution persists in speaker-indexed records (c--d) and an avatar-based replay (e).}
  \Description{Five panels show the VocalEyes experience: live AR captions with a named speaker profile and face marker, followed by a speaker list, a timestamped per-speaker transcript, and an avatar-based meeting replay.}
  \label{fig:teaser}
\end{teaserfigure}



\maketitle

\section{Introduction}\label{sec:intro}

In-person group conversations require people to continuously coordinate \emph{what} is being said with \emph{who} is saying it. As the number of participants increases, listeners must distribute their attention across speech, multiple conversational partners, and nonverbal cues, making group conversation cognitively demanding~\cite{zhou2026chatmuse}. Recent augmented and mixed reality systems have begun to support these interactions by presenting captions, contextual information, and conversational assistance directly within the shared physical space~\cite{marques2021conceptual,chatmuse}. However, for such information to remain useful, the interface must maintain its connection to the physical people who produced it. This connection becomes especially important when participants are unfamiliar with one another: knowing the words alone may not be sufficient when a listener needs to respond to a speaker, remember who contributed an idea, or later determine who made a statement.

Consider a project kickoff in which a specialist joins late. A fixed caption may show a proposed change, but the wearer may not know whether the specialist or another person proposed it. Looking away from the text to find a face can make the wearer miss the next sentence. A durable speaker name connects the current response to later review, while a transient location cue only guides attention during the visible turn.

This creates a \emph{speaker grounding} problem for AR conversation interfaces: how can the system establish and maintain the connection between an incoming utterance, a speaker's identity, and their physical presence in the environment? Prior work demonstrated the broader opportunity for MR to provide real-time support during in-person group conversations, while also showing that static information can require users to shift attention between an overlay and their conversational partners~\cite{zhou2026chatmuse}. Participants specifically expressed a preference for information that adapts to the person currently speaking. \vocaleyes~builds on this observation by addressing a prerequisite that speaker-aware interfaces often take for granted: how the system learns \emph{who} those conversational partners are in the first place.

\begin{figure*}[t]
\centering
\includegraphics[width=0.9\linewidth]{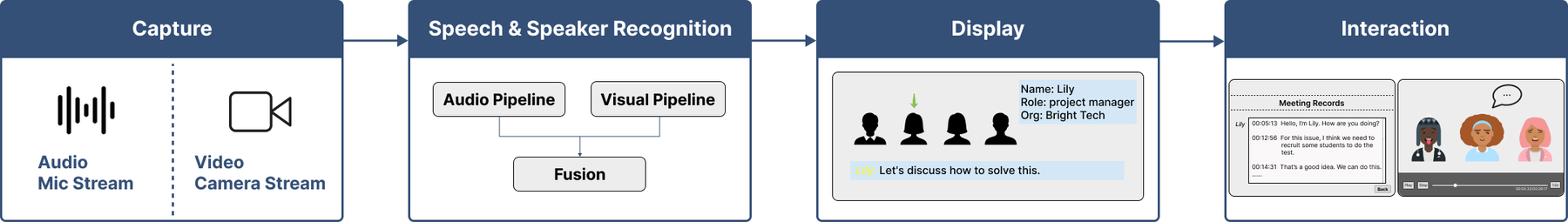}
\vspace{-3mm}
\caption{
    \textbf{An overview of our AR captioning system, \vocaleyes.}
}
\Description{A four-stage flow from audio and video capture, through parallel speech and visual processing, to a live AR display with speaker profile and captions, and then post-meeting records and avatar replay.}
\label{fig:system_overview}
\vspace{-4mm}
\end{figure*}

However, in everyday conversations, speaker identity is not always known in advance. Participants may be unfamiliar to the headset wearer, join after a conversation has begun, or never have provided enrollment samples. A speaker may therefore begin as an unknown voice and become identifiable only later through the conversation itself (\eg, through a natural self-introduction). Existing enrollment approaches may therefore require participants to pause the conversation to enroll, provide voice samples, or manually associate names with speakers, which conflicts with the lightweight and opportunistic nature of in-person AR interaction.

Existing captioning, diarization, and active-speaker systems each leave part of this problem unresolved. Diarization separates voices but commonly returns anonymous labels such as \emph{Speaker~1}~\cite{bain2023whisperx,AzureSpeechService}. Speaker-recognition systems can replace those labels with names, but they usually require a roster, manual annotation, or voice samples collected before the meeting~\cite{snyder2018x,dehak2010front}. Visual active-speaker detection can point toward an articulating face, but it does not establish that person's identity. These assumptions fit scheduled and instrumented meetings better than informal conversations or late arrivals~\cite{kushalnagar2019}. Thus, the remaining interaction challenge is not simply detecting \emph{who is speaking at a given moment}, but enabling speaker identity to be acquired, represented, and maintained as the conversation unfolds.

We investigate this challenge through \emph{in-conversation speaker registration}: an interaction paradigm in which an AR system incrementally constructs speaker identities from evidence that becomes available naturally during conversation, rather than requiring a separate enrollment stage. This introduces three interaction requirements. First, the system must represent speakers whose identities remain unresolved. Second, it must recognize identity-bearing conversational events and bind newly acquired semantic information to previously observed vocal evidence. Third, it must maintain this identity across subsequent utterances and connect it to the speaker's physical presence when visible.

We introduce \vocaleyes{}, a speaker-aware AR system that instantiates and evaluates this in-conversation registration paradigm. The system accumulates acoustic evidence under \textsc{Unknown Speaker}, treats a natural self-introduction as an identity-bearing event, and links the extracted name to that voice profile. This allows speaker identity to emerge during the conversation without interrupting participants for explicit enrollment. It then assigns complementary roles to two inference channels. Audio provides delayed but persistent identity and captions, while mouth motion provides a faster, view-dependent cue for the articulating face. The AR interface combines speaker-attributed captions, a stable profile card, and a spatial marker, and the meeting record preserves the resulting attributions (Figure~\ref{fig:system_overview}).

Importantly, VocalEyes does not aim to introduce a new speaker diarization or active-speaker detection model. Rather, it's a modular interactive system that integrates acoustic, visual, and semantic signals centered around speaker identity: identity is treated as a dynamic conversational state that may be unresolved initially, acquired opportunistically during interaction, and subsequently maintained across modalities. This modular architecture also allows us to examine the distinct contributions of speaker identity, spatial localization, and contextual profile information to the user's meeting experience.

We evaluated the interface through a controlled within-subjects study with $20$ participants. Five conditions isolated caption text, spatial location, audio-derived identity, and the profile card. In scripted meetings with trained confederates, \vocaleyes{} reached $88.0\%$ speaker-identification accuracy and increased participant tracking accuracy from $47.2\%$ with caption-only AR to $87.3\%$ with the complete interface. Participants also reported lower workload and higher usability. Beyond demonstrating the benefit of multimodal fusion, our component evaluation distinguishes how different forms of speaker information contribute to interaction: identity provides information about \emph{who} is speaking, spatial cues indicate \emph{where} the speaker is, and profile information provides contextual information about \emph{who that person is}. In this paper, we make three core contributions:

\begin{itemize}[noitemsep, topsep=0pt, leftmargin=2em] 
\item \textbf{An in-conversation speaker registration paradigm} that converts identity-bearing self-introductions into named, revisable voice profiles without a separate pre-meeting enrollment step. 
\item \textbf{A speaker-aware AR interaction architecture} that coordinates persistent speaker attribution, stable identity context, and view-dependent spatial guidance while retaining an explicit unknown-speaker state. 
\item \textbf{Empirical design knowledge for speaker-aware AR} from a controlled study ($N = 20$) that distinguishes the roles of identity, location, and profile context in speaker tracking, workload, and usability. 
\end{itemize}

\section{Related Work} \label{sec:rw}
Prior work provides three parts of speaker-aware captioning: AR interfaces that organize attention, audio systems that distinguish voices, and visual systems that localize articulation. We examine the assumptions behind each part and identify the interaction gap that appears when a meeting lacks a pre-enrolled speaker roster.

\subsection{Speaker Grounding in AR Conversation Interfaces} \label{sec:arconf}

AR conferencing systems coordinate distributed information with embodied people. Telepresence systems use avatars, gaze, spatial layout, and conversational formations to communicate presence and attention~\cite{abdullah2021videoconference,marques2021conceptual,piumsomboon2018mini,kendon1990conducting}. High-fidelity systems such as Holoportation reproduce remote bodies but require specialized capture hardware~\cite{ortescolano2016holoportation}. Lighter-weight systems instead use animations or directional cues to tell users where to look~\cite{kim2024dragon,li2024location,alghofaili2019lost,jing2015content}. Across these approaches, interface placement shapes how users divide attention between mediated content and the surrounding conversation.

AR captions make this grounding problem explicit because the caption and speaker occupy different locations. Spatial speech bubbles and see-through captions place an utterance near its visible source~\cite{peng2018speechbubbles,yamamoto2021see,zhang2022balloon}. However, a face-anchored caption can leave the field of view, compete with facial cues, or become crowded when multiple speakers sit close together. A fixed caption region preserves readability but weakens the visible link between the caption text and speaker. \vocaleyes{} treats these as complementary grounding functions: persistent attribution maintains \emph{who} produced the utterance, while spatial guidance indicates \emph{where} that person is located.

Prior speaker-aware caption interfaces primarily help users find the current source through placement, pointers, highlights, or shared transparent text~\cite{peng2018speechbubbles,kushalnagar2019,glasser2019mixed,yamamoto2021see,samaradivakara2024tailored}. These mechanisms strengthen current-turn orientation, but they generally assume that the system already knows how speakers should be represented. A longer-lived interaction requires the interface to preserve the same source across live captions, transcripts, and replay~\cite{kuichi2025minimates}. \vocaleyes{} addresses this continuity by allowing a speaker to remain unresolved initially, become identified during conversation, and then retain that identity in subsequent captions and meeting records.

Recent MR work also extends conversational support beyond caption presentation. ChatMuse analyzes verbal and nonverbal cues from a small group and proactively presents private guidance about the wearer's verbal and nonverbal behavior~\cite{zhou2026chatmuse}. It shows how MR can provide adaptive support during in-person conversation. \vocaleyes{} addresses a complementary problem: maintaining the connection between each utterance and the person who produced it as users shift attention during the meeting and later revisit the record.

Together, prior AR systems provide mechanisms for directing attention, identifying the current speaker, and supporting conversational participation. However, they do not establish an identity lifecycle for speakers who are initially unknown. In particular, they do not address how an interface can infer a speaker's identity from a natural conversational event, update representations of earlier and later utterances, and preserve that attribution in a post-meeting record. This is the interaction gap that motivates \vocaleyes{}.

\begin{table*}[t]
\centering
\small
\setlength{\tabcolsep}{2.5pt}
\caption{What representative AR conversation systems show about a speaker. ``Points now'' means that the interface directs attention to a current source. ``Learns name'' means that it acquires a speaker's name during the conversation. ``Names record'' means that stored utterances retain that name. Among these systems, only \vocaleyes{} reports all three functions.}
\label{tab:prior_work}
\begin{tabular}{@{}>{\raggedright\arraybackslash}p{38mm}>{\raggedright\arraybackslash}p{48mm}>{\centering\arraybackslash}p{17mm}>{\centering\arraybackslash}p{19mm}>{\centering\arraybackslash}p{20mm}@{}}
\toprule
\textbf{System} & \textbf{Live function} & \textbf{Points now} & \textbf{Learns name} & \textbf{Names record} \\
\midrule
SpeechBubbles~\cite{peng2018speechbubbles} & Spatial caption and off-screen cue & Yes & No & No \\
RTTD-ID~\cite{kushalnagar2019} & Pointer and pop-up & Yes & No & No \\
Glasser et al.~\cite{glasser2019mixed} & Light, glow, and pointer & Yes & No & No \\
See-Through Captions~\cite{yamamoto2021see} & Shared transparent captions & Yes & No & No \\
Tailored AR Captions~\cite{samaradivakara2024tailored} & Configurable caption placement & No & No & No \\
ChatMuse~\cite{zhou2026chatmuse} & Private conversational guidance & No & No & No \\
\vocaleyes{} & Named caption and face marker & Yes & Yes & Yes \\
\bottomrule
\end{tabular}
\end{table*}

\subsection{From Anonymous Diarization to In-Conversation Speaker Identity}\label{sec:realtimeSR}
Audio pipelines first decide when speech occurs, then transcribe and associate segments with voices. Voice activity detection ranges from statistical noise tracking to learned detectors~\cite{cohen2003noise,Silero_VAD}. Speaker-change systems can align acoustic changes with automatic speech recognition~(ASR) tokens~\cite{zheng2025scdiar}. Modern diarization tools make these capabilities increasingly available, but their output often remains a sequence of anonymous speaker clusters. A cluster supports turn segmentation; it does not by itself tell a caption reader which person produced the words.

Speaker verification can attach a stable identity when the system already has a voice sample. Embedding models represent short speech segments in a similarity space and compare them with enrolled profiles~\cite{dehak2010front,snyder2018x,jia2019transferlearningspeakerverification}. \vocaleyes{} uses Resemblyzer embeddings~\cite{2020resemblyzer} with Faster-Whisper transcription~\cite{radford2022whisper,bain2023whisperx,faster_whisper}.

This shift from pre-enrollment to in-conversation registration introduces interface obligations. Before a self-introduction, the system must avoid presenting a guessed name as fact. After registration, it must update later captions without hiding the transition from unknown to known. If the system assigns the wrong profile, users need a path to inspect and correct the association. We treat these states as part of the user experience rather than as internal diarization bookkeeping.

These models provide the acoustic and textual primitives that \vocaleyes{} uses. We propose to leverage existing interaction models for enrollment. Our system stores an unresolved voice until a name is available, detects a later self-introduction as an identity-bearing event, and associates the extracted name with the accumulated voice evidence. The result is a transition from anonymous acoustic attribution to a named speaker profile without requiring a separate enrollment stage.

\subsection{Visual Localization as a Complement to Speaker Identity}\label{sec:fusion}
Visual active-speaker detection estimates which visible face currently articulates speech. Optical flow and geometric lip tracking reduce computation but remain sensitive to pose, occlusion, and lighting~\cite{cutler2000look,matthews2002extraction}. Deep audio-visual models improve robustness by learning temporal correspondence across modalities, although many require greater computation than a mobile AR pipeline can sustain~\cite{chung2016out,petridis2018end,kundu2025efficient}. \vocaleyes{} follows geometric aperture-ratio methods~\cite{cech2016real} and computes mouth aspect ratio~(MAR) from MediaPipe landmarks~\cite{lugaresi2019mediapipe}.

Audio-visual fusion usually aims to improve a model-level prediction. Gating and cross-modal approaches combine acoustic and visual evidence to strengthen active-speaker decisions under noise and changing context~\cite{tao2018gating,alcazar2020active,vasireddy2024robust}. Datasets such as AVA-ActiveSpeaker and egocentric corpora support this algorithmic comparison~\cite{roth2020ava,jiang2022ego,sarkar2023self}. \vocaleyes{} instead treats audio and vision as complementary interface resources rather than as inputs to a single fused classifier: audio provides delayed but persistent identity and text, while vision provides faster, transient spatial guidance.

This separation lets us ask an HCI question that model-level fusion obscures. A single confidence score may improve classification, but it does not tell designers where, when, or for how long each cue should appear. \vocaleyes{} therefore evaluates identity, location, and profile context as user-visible information layers. The study examines what each layer contributes to speaker tracking and workload within a controlled meeting task.

Across these three areas, prior work provides important mechanisms for caption access, current-turn source finding, speaker separation, and conversational support (Table~\ref{tab:prior_work}). The unresolved issue is how these mechanisms should be coordinated when speaker identity is not known at the beginning of the interaction. \vocaleyes{} addresses this issue through an identity lifecycle that connects unresolved speech, in-conversation registration, named live attribution, and an attributed post-meeting record. Its novelty lies in this interaction lifecycle and in the user-visible coordination of existing speech and vision components.

\section{Design Rationale and Scope}\label{sec:designrationale}

We derive four design requirements from the speaker--utterance attribution task and the capability boundaries in prior work. We additionally draw on design considerations from ChatMuse's study of MR-supported in-person group conversations, which highlighted the need to account for both verbal and nonverbal conversational context while minimizing unnecessary visual distraction from real-time support~\cite{zhou2026chatmuse}. Prior spatial-caption work further motivates keeping text tied to its source while users shift attention~\cite{peng2018speechbubbles,yamamoto2021see,zhang2022balloon}. These objectives organize the system:

\noindent\fbox{\textbf{DO1}}~{\bf Keep speaker identity attached to the utterance.}
Caption readers need an attribution that survives shifts in gaze. A face marker can indicate where speech appears to originate, but the marker disappears when a user looks away or the speaker leaves the field of view. This follows the broader design consideration identified in ChatMuse that MR conversational support must remain interpretable as users divide attention between digital information and multiple conversational partners~\cite{zhou2026chatmuse}. \vocaleyes{} therefore prefixes each caption with the resolved speaker name. This placement preserves the relationship between speaker and text at the point where users read it and carries the same relationship into the meeting record.

\noindent\fbox{\textbf{DO2}}~{\bf Turn natural introductions into registration events.}
Poorly timed MR support can interrupt a live conversation~\cite{zhou2026chatmuse}. A separate enrollment task creates the same risk. \vocaleyes{} gathers voice evidence before it knows a name, listens for a self-introduction, and connects the extracted identity to the temporary voice buffer. Until that event occurs, the interface labels the source \textsc{Unknown Speaker}. This explicit state lets the system defer attribution rather than guess.

\noindent\fbox{\textbf{DO3}}~{\bf Assign identity and location to different interface layers.}
Audio and vision produce different evidence on different schedules. ChatMuse similarly found that supporting in-person group conversation requires jointly considering verbal and nonverbal context rather than relying on either source alone~\cite{zhou2026chatmuse}. In \vocaleyes{}, we use these complementary signals for different interface functions rather than requiring them to communicate the same information. The audio path can associate an utterance with a named voice profile, but it must wait for sufficient speech and transcription. The visual path can react to mouth motion within a few frames, but it identifies an articulating face rather than a person. \vocaleyes{} maps these outputs to separate objects: the caption carries audio-derived identity, and a marker carries visual location. The interface hides the marker when vision lacks an active face and retains \textsc{Unknown Speaker} when audio lacks an identity.

\noindent\fbox{\textbf{DO4}}~{\bf Separate immediate attribution from persistent identity context.}
A speaker name answers the immediate question of who produced an utterance. Role and organization answer a different question: how an unfamiliar participant relates to the meeting. ChatMuse identified the presentation of real-time MR information as a design tradeoff: conversational support should provide useful context without introducing unnecessary visual distraction~\cite{zhou2026chatmuse}. Placing every field inside a moving caption or face anchor would increase visual motion and compete with speech content. \vocaleyes{} therefore places the name, role, and organization in a fixed top-right profile card. The stable position supports repeated reference, while the bottom caption remains concise. The specific metadata fields are auxiliary to the core design (\ie, we selected name, role, and organization because they were relevant to our study's meeting scenarios), and other applications could substitute different contextual information. Bundling the three fields was a prototype design choice. The study compares the complete card with no card; it does not isolate the value of each field.

\noindent{\bf Study operationalization.}
The five conditions vary the information available to participants across four interface layers. \emph{Baseline} provides caption text alone. \emph{Visual Only} adds location. \emph{Audio Only} adds identity and the profile card. \emph{Combined Cues w/o Profile} combines identity with location but removes the card. \vocaleyes{} presents all four layers. These contrasts test whether users benefit from knowing where speech occurs, who produced it, and which stable context accompanies that identity.

\begin{figure*}[t]
\centering
\includegraphics[width=\textwidth]{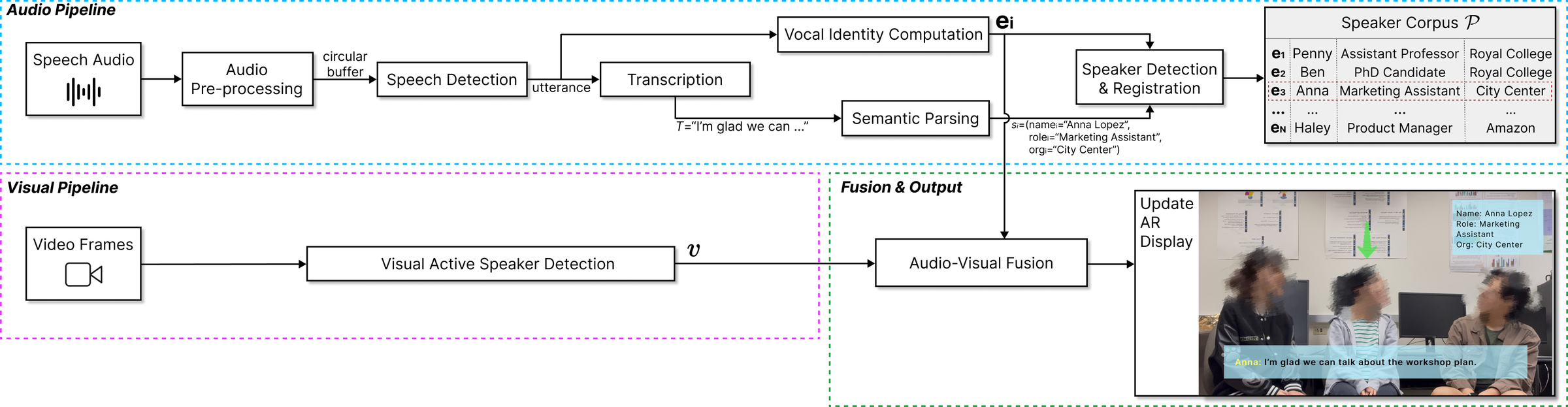}
\vspace{-7mm}
\caption{Overview of the real-time speaker-aware captioning pipeline. Independent audio and video services process the two streams. The audio service detects speech, extracts a speaker embedding ($\mathbf{e}_i$), transcribes text ($T$), and derives semantic information ($s_i$). The visual service computes mouth-motion confidence $v$. The AR interface maps audio to identity and captions and maps vision to a spatial cue; the prototype does not bind voices to face identities.}
\Description{The upper audio branch transforms speech through preprocessing, detection, transcription, vocal-identity computation, and semantic parsing into a speaker corpus. The lower video branch detects visual speech activity. Their outputs drive the AR profile, caption, and face marker.}
\label{fig:speaker_recog}
\vspace{-3mm}
\end{figure*}

\section{System Overview}

Figure~\ref{fig:system_overview} summarizes four stages. A microphone and camera first capture the meeting. The audio service then segments speech, transcribes it, and registers identity from introductions (DO2). The visual service separately tracks faces and mouth motion (DO3). The Unity client keeps names with captions (DO1), places role and organization in a stable card (DO4), and shows a transient face marker (DO3). Finally, \vocaleyes{} retains speaker attribution in the meeting record (DO1).

\section{Speech and Speaker Recognition}\label{sec:speachrecog}
The recognition pipeline turns audio and video into two interface outputs: named captions and a face-localized activity cue (Figure~\ref{fig:speaker_recog}). This section describes audio processing, in-conversation registration, visual detection, and interface-level coordination.

\subsection{Audio Pre-Processing} \label{sec:preprocess}
We convert continuous microphone input into utterances for transcription and speaker identification (Supplementary Figure~S1). A \emph{frame} contains $200$\,ms of audio. A \emph{speech segment} joins consecutive speech frames until silence or a forced flush. An \emph{utterance} passes the minimum-duration filter.

\ssection{Audio Capture.} A microphone beside the wearer samples audio at $16$\,kHz.

\ssection{Segmentation.}
We read audio in $200$\,ms frames. Pilot comparisons showed that $50$\,ms frames reacted to noise, while $500$\,ms frames merged turns and missed short segments (Supplementary Figure~S2). The $200$\,ms setting kept front-end delay below one frame while preserving temporal context. ASR dominates the longer end-to-end delay (Section~\ref{sec:ardesign}).

\ssection{Buffering.}
The pipeline appends each speech frame to an $8$-second circular buffer. Three silent frames ($0.6$\,s) close the segment. The pipeline also flushes a full buffer to bound latency during long turns.

\subsection{Speech Detection} \label{sec:speech_detection}
The voice activity detector compares each frame's peak amplitude with $\tau_{\text{vad}} = 0.03$ on a normalized $[0,1]$ scale. Frames below the threshold increment the silence counter; three consecutive silent frames close the segment. The pipeline discards segments shorter than $1.5$\,s and sends longer utterances to transcription and speaker identification. Pilot recordings from quiet meeting rooms guided the threshold, so non-stationary noise can trigger false detections (Section~\ref{sec:dcs}).

\subsection{Transcription} \label{sec:transcription}
Faster-Whisper transcribes each utterance into $T$~\cite{faster_whisper}. We run Whisper-small with \textsc{int8} quantization on the companion workstation. The interface displays $T$, and the identity parser extracts self-introduction fields from it (Section~\ref{sec:speakeriden}).
\subsection{Speaker Identification} \label{sec:speakeriden}
The identification pipeline links acoustic evidence to identity fields from self-introductions. It maintains a dynamic corpus $\mathcal{P} = \{(\mathbf{e}_i, \mathbf{s}_i)\}_{i=1}^N$, where profile $i$ stores a voice embedding $\mathbf{e}_i$ and $\mathbf{s}_i = (\text{name}_i, \text{role}_i, \text{org}_i)$.

\subsubsection{Extract Vocal Identity}
Resemblyzer converts each utterance into a $256$-dimensional speaker embedding~\cite{2020resemblyzer}. The CPU processes one embedding in about $40$\,ms. At startup, a background thread warms the encoder with silence to avoid first-use delay.

\ssection{Vocal Similarity.} We use cosine similarity to measure the similarity between two vocal embeddings:
{\setlength{\abovedisplayskip}{0.5pt}
\setlength{\abovedisplayshortskip}{0.5pt}
\setlength{\belowdisplayskip}{0.5pt}
\setlength{\belowdisplayshortskip}{0.5pt}
\begin{equation}
\label{eq:vocal_sim}
\varphi_{\text{vocal}}(\mathbf{e}_i, \mathbf{e}_j) \;=\; \frac{\mathbf{e}_i\cdot \mathbf{e}_j}{\lVert \mathbf{e}_i\rVert \cdot \lVert \mathbf{e}_j\rVert}
\end{equation}
where $\mathbf{e}_i$ and $\mathbf{e}_j$ represent utterances $i$ and $j$.
}
\subsubsection{Speaker Change Detection.} \label{sec:speaker_change}
The system triggers identification when vocal similarity signals a speaker change. A sliding window holds five $200$\,ms speech frames. At each step, Resemblyzer embeds the window and compares it with the last confirmed speaker embedding $\mathbf{e}_k$. The system detects a change when $\varphi_{\text{vocal}}(\mathbf{e}_i, \mathbf{e}_k) < \theta_{\text{vocal}}$, with $\theta_{\text{vocal}} = 0.72$. This lower threshold flags a potential boundary before the matching rules in Section~\ref{sec:speaker_detection}. Supplementary Figure~S4 shows one representative transition. During stable turns, a weighted average ($\gamma = 0.3$) adapts $\mathbf{e}_k$ to vocal variation.

At a detected change, the system sends earlier frames to transcription and speaker identification, then starts the next buffer with the remaining frames. If the first window triggers a change, the system sends the full window and clears the buffer. An $8$-second force flush bounds delay during monologues and resets the diarization state.

\subsubsection{Semantic Parsing}
Semantic parsing extracts $\mathbf{s}_i = \{name_i, role_i, org_i\}$ from a speaker's self-introduction. An LLM first returns these fields as structured JSON. If that call fails, spaCy named-entity recognition and regular expressions recover available fields~\cite{spacy2}.

\ssection{LLM-Based Parsing.\label{sec:llm-parsing}} We prompt GPT-3.5-turbo to extract $name_i$, $role_i$, and $org_i$ from the accumulated transcript~\cite{openai2022chatgpt}. For example, ``Hi, my name is Lily Chen and I'm a project manager at Bright Future Tech Company'' yields \texttt{\{"name": "Lily Chen", "role": "Project Manager", "org": "Bright Future Tech Company"\}}. The parser handles introduction forms that fixed patterns miss.

The service appends consecutive utterances from the same speaker before each query. After four failed extractions, it pauses queries until eight more utterances arrive. This cooldown limits repeated calls on speech without identity content.

\ssection{Rule-Based Parsing.\label{sec:rule-parsing}} An empty, incomplete, or invalid LLM response activates the fallback. It combines spaCy PERSON and ORG entities with the regular expressions in Supplementary Table~S1. Because VAD removes segments shorter than $1.5$\,s, the parser never receives brief interjections. This filter reduces spurious metadata but also drops genuine backchannels (Section~\ref{sec:dcs}).

\subsubsection{Speaker Detection and Registration}\label{sec:speaker_detection}
An unrecognized speaker may talk several times before introducing themself. To preserve that acoustic evidence, the system matches each incoming embedding $\mathbf{e}_i$ against registered profiles and temporary buffers:
{\setlength{\abovedisplayskip}{0.5pt}
\setlength{\abovedisplayshortskip}{0.5pt}
\setlength{\belowdisplayskip}{0.5pt}
\setlength{\belowdisplayshortskip}{0.5pt}
\begin{equation}
\label{eq:eq4}
\begin{cases}
\text{Update speaker embedding} & \text{if } \exists\, j \text{ s.t. } \varphi_\text{vocal}(\mathbf{e}_i, \mathbf{e}_j) \geq \theta_\text{corpus}, \\
\text{Update temp buffer} & \text{else if } \exists\, t \text{ s.t. } \varphi_\text{vocal}(\mathbf{e}_i, \bar{\mathbf{e}}_t) \geq \theta_\text{temp}, \\
\text{Open new temp buffer} & \text{otherwise.}
\end{cases}
\end{equation}
}where $j$ indexes profiles in $\mathcal{P}$ and $\bar{\mathbf{e}}_t$ denotes temporary buffer $t$'s centroid. Each buffer stores at most $M = 15$ embeddings. We set $\theta_\text{corpus} = 0.85$ and $\theta_\text{temp} = 0.75$. Because the acceptance regions overlap, the system checks registered profiles first. Similarities between the thresholds remain unattributed while the system gathers evidence. Supplementary Figure~S7 shows the pilot distributions.

\ssection{Unregistered Speaker Set.} 
If no corpus profile matches, the system adds $\mathbf{e}_i$ to the closest eligible temporary buffer:
{\setlength{\abovedisplayskip}{0.5pt}
\setlength{\abovedisplayshortskip}{0.5pt}
\setlength{\belowdisplayskip}{0.5pt}
\setlength{\belowdisplayshortskip}{0.5pt}
\begin{equation}
    \mathbf{E}_t \leftarrow 
    \begin{cases} 
        \big(\mathbf{E}_{t,1},\ \ldots,\ \mathbf{E}_{t,|\mathbf{E}_t|},\ \mathbf{e}_i\big) & \text{if } |\mathbf{E}_t| < M, \\[2pt]
        \big(\mathbf{E}_{t,2},\ \ldots,\ \mathbf{E}_{t,M},\ \mathbf{e}_i\big) & \text{otherwise,}
    \end{cases}
\end{equation}
}where $M = 15$ bounds memory and keeps the centroid focused on recent speech.
 
\ssection{Update Speaker Embedding.} When the corpus contains a match, the system selects $j^*$:
{\setlength{\abovedisplayskip}{0.5pt}
\setlength{\abovedisplayshortskip}{0.5pt}
\setlength{\belowdisplayskip}{0.5pt}
\setlength{\belowdisplayshortskip}{0.5pt}
\begin{equation}
    \label{eq:best_match}
    j^* = \arg\max_{j} \varphi_{\text{vocal}}(\mathbf{e}_i,\mathbf{e}_j)
\end{equation}
It then updates the stored embedding through an exponential moving average:
\begin{align}
\label{eq:update_emb}
\mathbf{e}_{j^*} &\leftarrow (1-\beta)\,\mathbf{e}_{j^*} \;+\; \beta\,\mathbf{e}_i
\end{align}
}where $\beta = 0.35$. The service batches disk writes to reduce I/O.

\ssection{Speaker Registration in Corpus.} 
The service parses transcripts associated with temporary buffers. When parsing yields a name, it inserts the buffer centroid and $s_i = (name_i, role_i, org_i)$ into $\mathcal{P}$. This promotion preserves acoustic evidence from before the introduction. The service also relabels earlier utterances that used the temporary identifier.

\subsection{Visual Active Speaker Detection} \label{sec:visual} 
\begin{figure}[t]
  \centering
    \includegraphics[width=\linewidth]{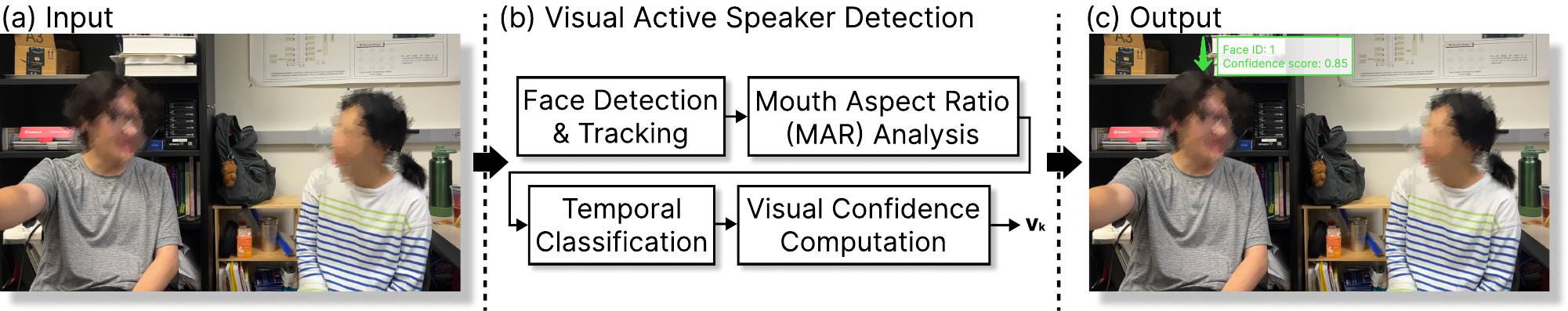}
    \vspace{-7mm}
  \caption{Visual Active Speaker Detection Pipeline}
  \Description{A flow diagram shows video frames entering face detection and tracking, facial landmark extraction, mouth-aspect-ratio calculation, temporal mouth-motion analysis, confidence scoring, and active-speaker selection.}
  \label{fig:fig10}
      \vspace{-5mm}
\end{figure}

The visual pipeline provides a fast, audio-independent location cue (Figure~\ref{fig:fig10}). It detects faces, tracks mouth motion, and selects the strongest articulating face on a CPU.

\ssection{Face Detection and Tracking.} 
MediaPipe FaceDetection localizes up to six faces, and Face Mesh returns $468$ landmarks per face~\cite{lugaresi2019mediapipe}. The service detects faces every second frame, updates landmarks every frame, and rejects boxes smaller than $50$\,px. It assigns each landmark set to the nearest box centroid.

Greedy Intersection over Union~(IoU) matching preserves face tracks across frames. The service accepts matches above $0.3$ and stores each track's ID, box, landmarks, last-seen frame, and absence count. It releases tracks after $90$ absent frames.

\ssection{Mouth Aspect Ratio Analysis.} We measure mouth aspect ratio~(MAR) from six upper and lower lip landmarks~\cite{matthews2002extraction}. The service averages the left, center, and right vertical lip distances and divides by mouth width. This normalization reduces distance effects. Static mouth openness cannot reliably indicate speech, so the detector uses MAR changes over time.
Supplementary Figure~S3 illustrates the computation $\text{MAR} = \frac{d_\text{center} + d_\text{left} + d_\text{right}}{3 \cdot d_{\text{mouth}}}$ from three vertical lip distances and mouth width.

\ssection{Mouth-Based Speech Detection.} 
For each face, a ten-frame window supplies three motion features: consecutive-frame change, window variance, and the largest change over five frames. At $10$\,FPS, the window spans about $1$\,s. The detector marks speech when at least two features exceed their thresholds: $0.04$ for change, $0.002$ for variance, and $0.032$ for recent peak. This majority rule rejects static open mouths and many isolated movements.

A counter stabilizes the decision. It rises when the majority rule holds, falls otherwise, and remains within $[-5,20]$. The detector activates a face at $6$ and deactivates it at $-3$. This hysteresis reduces flicker.

\ssection{Visual Confidence Score.} We compute a continuous visual confidence score $v \in [0,1]$ as a weighted combination of temporal stability, motion intensity, and instantaneous activity:
{\setlength{\abovedisplayskip}{0.5pt}
\setlength{\abovedisplayshortskip}{0.5pt}
\setlength{\belowdisplayskip}{0.5pt}
\setlength{\belowdisplayshortskip}{0.5pt}
\begin{equation}
\label{eq:visual_conf}
v = 
0.4 \cdot \min\!\left(1, \frac{c_s}{15}\right)
+ 0.3 \cdot \min\!\left(1, \frac{\sigma_s}{3\,\theta_{\text{var}}}\right)
+ 0.3 \cdot \min\!\left(1, \frac{\Delta_s}{2\,\theta_{\text{rate}}}\right)
\end{equation}}
where $c_s$ denotes the counter, $\sigma_s$ denotes ten-frame MAR variance, and $\Delta_s$ denotes instantaneous MAR change. The service sets $v=0$ for inactive, absent, or occluded faces and evaluates Equation~\ref{eq:visual_conf} only when $c_s \geq 6$. Thus, $v$ measures confidence in articulation, not facial identity.

\subsection{Combining the Two Channels} \label{sec:avfusion}
The channels answer different questions. Audio identifies \emph{who spoke} and \emph{what they said} after the system closes and processes an utterance. Vision identifies \emph{which visible face is articulating} within a few frames but cannot name that face.

\vocaleyes{} therefore keeps $\varphi_{\text{vocal}}$ and $v$ separate. The speaker corpus indexes acoustic identities, while the tracker assigns local face IDs. Without voice-to-face binding, a combined score could raise one person's acoustic confidence because another visible face moved. Section~\ref{sec:dcs} discusses this limitation. Instead, the interface composes the channels by assigning each one a supported display role (Section~\ref{sec:ardesign}):
\begin{itemize}[noitemsep, topsep=2pt, leftmargin=2em]
  \item The visual channel places the spatial marker above the face with the highest active $v$ and maps $v$ to opacity.
  \item The acoustic channel fills the profile card and caption.
  \item The interface hides the marker when vision finds no active face. It retains the marker with an \textsc{Unknown Speaker} label when audio has not resolved identity.
\end{itemize}

This composition lets vision answer ``where'' quickly and audio answer ``who'' and ``what'' later. Section~\ref{sec:ardesign} quantifies the latency difference.

\section{The Real-Time AR Captioning System}\label{sec:ardesign}
This section describes the deployment architecture, live visualization, post-meeting tools, and system-level tradeoffs.

\ssection{Deployment Architecture.} Three processes communicate over a local network. A companion workstation runs separate vision and audio services. A Magic Leap~2 Unity client polls visual state at $10$\,Hz and audio state at $5$\,Hz, then handles rendering and input. This split keeps sustained model inference off the headset and supports independent component ablations.

\subsection{AR Visualization of Speaker and Speech}
\vocaleyes{} presents three coordinated layers (Figure~\ref{fig:teaser}a--b). A top-right card shows $name_i$ and optional $role_i$ and $org_i$. Its fixed position keeps multi-line text legible as faces move. Before registration, the card shows \textsc{Unknown Speaker}. \setlength{\columnsep}{2pt}

\begin{figure}
  \centering
  \includegraphics[width=0.7\linewidth]{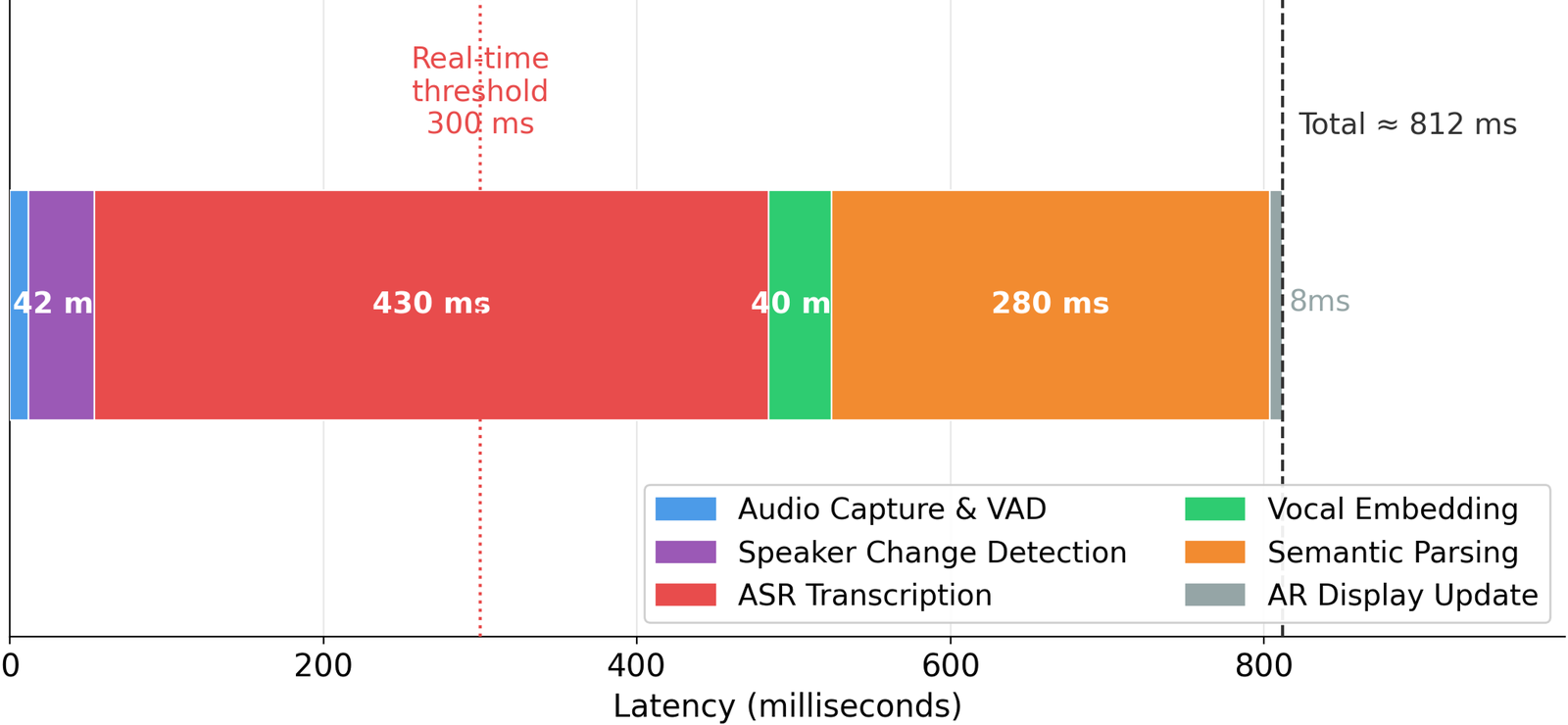}
  \caption{Pilot CPU processing-time profile after endpointing releases an identity-bearing audio segment. The reported $\approx812$\,ms excludes speech duration and the silence boundary, so it does not represent speech-onset-to-display latency. The $300$\,ms marker denotes the front-end engineering target in Supplementary Figure~S2, not a human perceptual threshold or a complete-pipeline target.
  }
  \Description{A stacked processing-time profile divides approximately 812 milliseconds after segment release among transcription, embedding, identity matching, semantic parsing, and interface update stages; a separate 300 millisecond marker denotes only the front-end target.}
  \label{fig:latency}
  \vspace{-5mm}
\end{figure}

A green marker points to the face with the highest visual confidence $v$; the interface hides it when vision finds no articulating face. At the bottom, the caption prefixes $T$ with the resolved name, so users can retain attribution without looking at the speaker.

\subsection{Post-Meeting Reflection}

During the meeting, \vocaleyes{} appends each non-empty utterance, timestamp, and available identity fields to a per-speaker record. It stores unresolved utterances under temporary IDs and relabels them after registration. At the end, it writes the roster, transcripts, and summary statistics to one session file.

The review interface lists speakers by contribution count and opens a timestamped transcript for the selected person (Figure~\ref{fig:teaser}c--d). Users select and scroll with the Magic Leap~2 controller. A second view replays utterances as speech bubbles above color-coded avatars (Figure~\ref{fig:teaser}e), following prior avatar-based AR work~\cite{kuichi2025minimates}. Because the record omits face positions, the replay reconstructs speaking order rather than spatial arrangement.

\subsection{System Design Considerations}

\ssection{Privacy.} The workstation processes raw audio and video within the headset's local network~\cite{rajaram2025privacy,rajaram2023multiar}. Semantic parsing sends transcript text, but not audio, images, or embeddings, to a hosted language model. A local language model could remove that external transfer.

The service discards raw audio after transcription and embedding and never stores face images or landmarks. It retains $256$-dimensional speaker embeddings and timestamped transcripts with identity fields. Because researchers treat speaker embeddings as linkable biometric identifiers, the interface provides meeting-record deletion~\cite{nautsch2019preserving,tomashenko2022voiceprivacy}.

\ssection{Interactivity.}
The two channels expose different delays. The visual path completes in about $70$--$200$\,ms and usually takes $100$--$150$\,ms on our local network. Detection and tracking consume $20$--$40$\,ms per frame for five or six faces at $640\times480$; encoding, transport, polling, and rendering account for the remainder.

The acoustic path takes about $1.5$--$4$\,s from speech onset to caption display. Segment duration and the $0.6$\,s endpoint precede transcription, embedding, matching, and display. Figure~\ref{fig:latency} reports about $812$\,ms after segment release for an identity-bearing segment. Semantic parsing runs only when the service creates or enriches a profile. This latency difference assigns rapid spatial guidance to vision and delayed identity and text to audio.

\ssection{Parameter Selection.} Three pilot participants completed scripted two- and three-speaker conversations. We applied the deployed VAD settings and removed segments shorter than $1.5$\,s. Pairwise similarities across speakers estimated $p(\varphi_\text{vocal}\mid\emph{unknown})$, while same-speaker pairs estimated $p(\varphi_\text{vocal}\mid\emph{corpus})$. Comparisons between each embedding and a centroid of $15$ same-speaker embeddings estimated $p(\varphi_\text{vocal}\mid\emph{temp})$. These distributions guided $\theta_\text{temp} = 0.75$ and $\theta_\text{corpus} = 0.85$ (Supplementary Figure~S7).

\section{User Study}\label{sec:us}
We conducted a within-subjects study to test how the interface layers affect speaker tracking, workload, and usability in controlled multi-speaker meetings. After the five ablation conditions, participants explored the post-meeting tools and provided formative feedback.

We investigate four research questions:

\noindent\textbf{RQ1.} What accuracy and processing time do the enabled speaker-identification and localization pipelines achieve in the controlled meetings?

\noindent\textbf{RQ2.} How do identity, location, and profile context affect participants' speaker-tracking accuracy?

\noindent\textbf{RQ3.} How do these information layers affect workload?

\noindent\textbf{RQ4.} How do participants rate the five interfaces and respond to the complete system's live and post-meeting features?

\ssection{Participants.}
We recruited $20$ participants ($9$ female, $11$ male) ages $20$--$38$ ($M = 26.5$, $SD = 5.7$) through campus bulletin boards and department mailing lists. All participants reported normal or corrected-to-normal vision. Their prior AR or VR experience ranged from zero to four years ($M = 1.2$, $SD = 1.1$), and none had used Magic Leap~2. The Institutional Review Board approved the study.
\begin{table}[t]
\centering
\footnotesize
\setlength{\tabcolsep}{4pt}
\caption{\label{tab:conditions}Study conditions and their active user-visible information. \checkmark~= active; \textemdash~= suppressed. ``Audio identity'' denotes the audio-derived name used for attribution; ``profile card'' denotes the fixed name/role/organization display.}
\vspace{-4mm}
\begin{tabularx}{\linewidth}{@{} p{28mm} *{4}{>{\centering\arraybackslash}X} @{}}
\toprule
  & \textbf{Caption}
  & \textbf{Spatial Cue}
  & \textbf{Audio Identity}
  & \textbf{Profile Card} \\
\midrule
\emph{Baseline (Caption Only)} 
  & \checkmark & --- & --- & --- \\

\emph{Visual Only} 
  & \checkmark & \checkmark & --- & --- \\

\emph{Audio Only} 
  & \checkmark & --- & \checkmark & \checkmark \\

\emph{Combined Cues w/o Profile} 
  & \checkmark & \checkmark & \checkmark & --- \\

\vocaleyes 
  & \checkmark & \checkmark & \checkmark & \checkmark \\
\bottomrule
\end{tabularx}
\vspace{-5mm}
\end{table}

\ssection{Apparatus.}
Participants wore a Magic Leap~2 running our Unity application. Over Wi-Fi, the headset connected to a Python~3.10 workstation that ran MediaPipe face tracking and MAR analysis~\cite{lugaresi2019mediapipe}, Faster-Whisper transcription~\cite{faster_whisper}, and Resemblyzer identification~\cite{2020resemblyzer}. The system sent $640\times480$ JPEG frames at $10$\,FPS and sampled workstation-microphone audio at $16$\,kHz. An experiment logger stored outputs by participant and condition. A wall-mounted camera recorded ground truth.

\ssection{Study Design and Procedure.}
Each participant completed the five conditions in Table~\ref{tab:conditions}:
\begin{itemize}[noitemsep, topsep=0pt, leftmargin=2em]
\setlength{\topsep}{0pt}
\setlength{\partopsep}{0pt}
\setlength{\itemsep}{0pt}
\setlength{\parskip}{0pt}
\setlength{\parsep}{0pt}
  \item \emph{Baseline (Caption Only)} showed captions without identity or location cues.
  \item \emph{Visual Only} added the MAR-driven location marker.
  \item \emph{Audio Only} added names in captions and the profile card.
  \item \emph{Combined Cues w/o Profile} combined names and the location marker but hid the card.
  \item \vocaleyes{} showed captions, names, the marker, and the card.
\end{itemize}

All $20$ participants wore the headset and completed every condition. Five counterbalancing groups assigned five condition orders. Each condition lasted $5$--$7$ minutes, followed by a five-minute break and questionnaire. Participants then explored the post-meeting interface and completed a $10$--$15$ minute semi-structured interview about utility, trust, and adoption. During the interview, participants could also enter written suggestions in an open-response field.

\ssection{Experimental Task.}
We used structured, scripted discussions to maintain comparable speaking patterns, turn distributions, and introduction events across conditions. This control was particularly important for the within-subject ablation, where differences in conversational dynamics could otherwise confound comparisons between speaker-identification cues. The scenarios were designed to approximate small-group discussions, but were not intended to reproduce the variability of unconstrained meetings.

Each participant met with the same three trained confederates. In the four identity-enabled conditions, two confederates began the meeting and the third joined midway. A script held turn frequency at about $8$--$10$ turns per speaker. It also kept speech length, topic transitions, and the introduction event similar across conditions. The confederates remained in fixed seats, and the room stayed quiet.

Before each participant's session, the service cleared its speaker corpus and registered the three confederates anew during the meeting.

\ssection{Measures and Analysis.}
We matched each research question to a distinct set of measures. For RQ1, we compared the pipeline's utterance-level speaker decision with annotations from the wall-mounted camera and logged processing latency after endpointing released a segment. For RQ2, randomized in-meeting probes asked participants to name the current speaker. For RQ3, participants completed NASA-TLX after every condition~\cite{hart1988nasa}. For RQ4, they completed the System Usability Scale~(SUS), rated individual \vocaleyes{} features, and discussed their experience in the final interview~\cite{brooke1996sus}.

We used each participant as the unit of analysis. Friedman tests examined condition effects, and Wilcoxon signed-rank tests supported post-hoc pairwise comparisons. We applied a Bonferroni correction within each family of comparisons and report the adjusted decision thresholds with the results. All $20$ participants completed every condition, so the analysis included no condition-level missing data or participant exclusions. We reproduce selected written suggestions verbatim as illustrative accounts and reserve all claims about themes, prevalence, and condition effects for the behavioral and questionnaire measures.

\begin{figure}[t]
\centering
\begin{subfigure}[t]{0.478\linewidth}
    \centering
    \includegraphics[width=\linewidth]{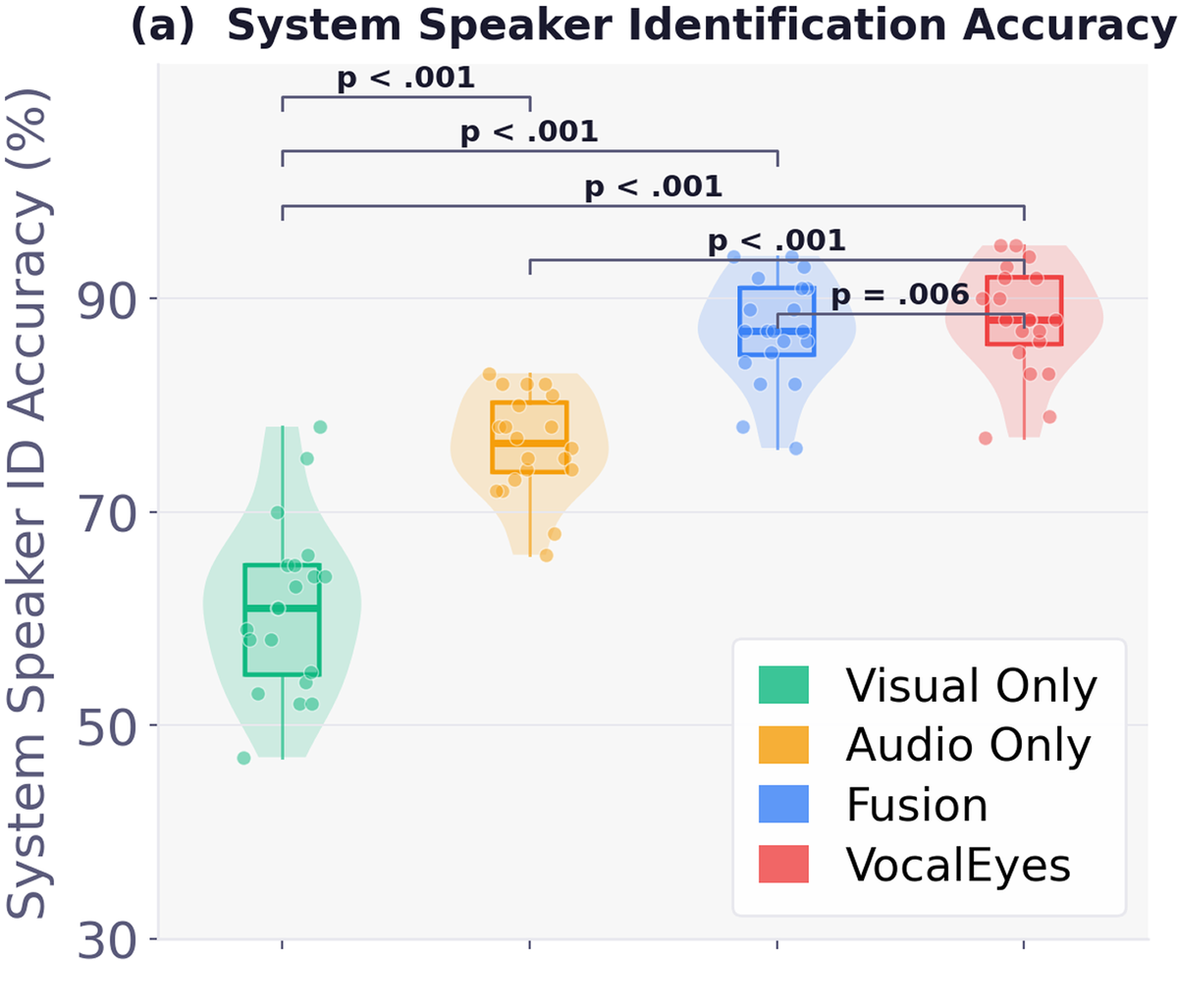}
\end{subfigure}
\hfill
\begin{subfigure}[t]{0.511\linewidth}
    \centering
    \includegraphics[width=\linewidth]{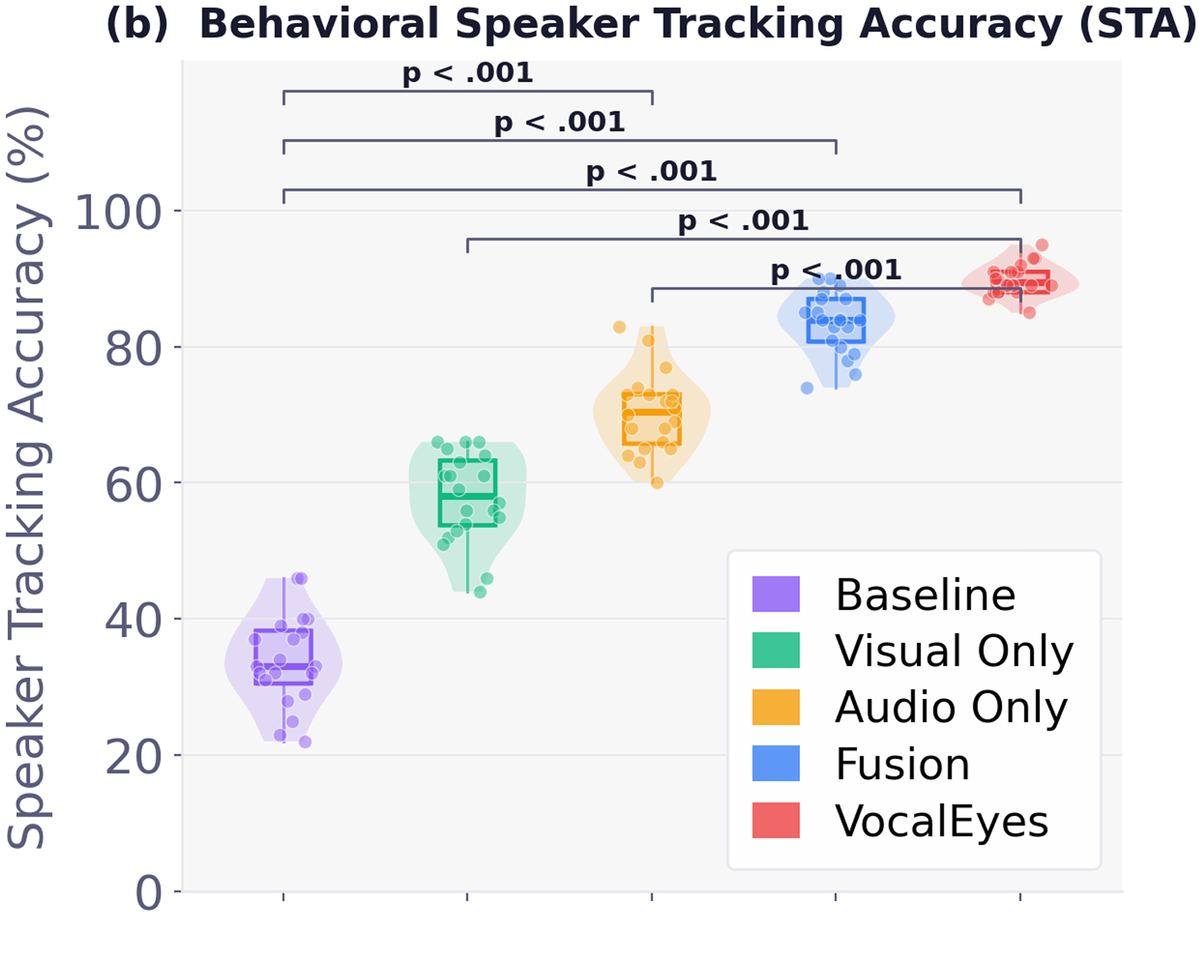}
\end{subfigure}
\vspace{-4mm}
\caption{(a) System speaker-identification accuracy and (b) participant speaker-tracking accuracy across conditions. Violins show full distributions, boxes indicate interquartile range (IQR), and dots represent individual participants. Panel~(a) displays the observed range from $30\%$ to $100\%$ rather than a zero-based axis. The figure legend's abbreviated ``Fusion'' label corresponds to \emph{Combined Cues w/o Profile} in the text.}
\Description{Two violin-and-box plots compare study conditions. VocalEyes leads system accuracy, followed by Combined Cues without Profile, Audio Only, and Visual Only. VocalEyes also leads participant tracking, while the caption-only baseline ranks last.}
\label{fig:ablation}
\vspace{-6mm}
\end{figure}

\subsection{Results}\label{sec:rs}
We organize the findings by research question. The first two findings distinguish technical speaker decisions from participants' ability to track a conversation. The final two findings connect workload and usability to participants' accounts of the interface.

\subsubsection{Combined Cues Produced the Highest System Accuracy (RQ1)}
The enabled pipelines differed in utterance-level speaker accuracy. Figure~\ref{fig:ablation}a summarizes the four conditions; \emph{Baseline} made no speaker decision and therefore does not appear in this comparison. A Friedman test found a condition effect ($p < .001$). Bonferroni-corrected Wilcoxon tests across six pairs used $\alpha_\text{Bonf} = .0083$. \vocaleyes{} reached the highest accuracy ($M = 88.0\%$, $SD = 2.87$) and outperformed \emph{Combined Cues w/o Profile} ($M = 87.0\%$, $SD = 5.09$; $p = .006$), \emph{Audio Only} ($M = 76.3\%$, $SD = 4.86$; $p < .001$), and \emph{Visual Only} ($M = 61.0\%$, $SD = 8.07$; $p < .001$). \emph{Combined Cues w/o Profile} outperformed both single-channel conditions, and \emph{Audio Only} outperformed \emph{Visual Only}. The two combined conditions used the same recognition pipeline in separate meeting trials, so we interpret their observed difference as trial variation rather than an effect of profile-card visibility on recognition. 

The channels also exposed different response times. Post-segment processing averaged $275$\,ms ($SD = 40$) for \emph{Visual Only}, $780$\,ms ($SD = 75$) for \emph{Combined Cues w/o Profile}, $812$\,ms ($SD = 70$) for \vocaleyes{}, and $840$\,ms ($SD = 83$) for \emph{Audio Only}. These measurements start after endpointing and exclude both speech accumulation and the silence boundary. They therefore describe processing cost rather than complete speech-onset-to-display latency. During the controlled late-entry event, the audio service registered Lulu from her first identity-bearing utterance and used the resulting profile without pausing the meeting. Supplementary Figure~S6 illustrates one registration event and the subsequent attributions.

\subsubsection{Identity Drove Participant Speaker Tracking (RQ2)}
Participants tracked speakers most accurately when captions carried audio-derived identity. Figure~\ref{fig:ablation}b summarizes the randomized in-meeting probes across all five conditions. Bonferroni-corrected Wilcoxon tests over ten pairs used $\alpha_\text{Bonf} = .005$. \vocaleyes{} ($M = 87.3\%$, $SD = 8.74$) outperformed \emph{Baseline} ($M = 47.2\%$, $SD = 19.8$; $p < .001$), \emph{Visual Only} ($M = 62.0\%$, $SD = 12.72$; $p < .001$), and \emph{Audio Only} ($M = 71.5\%$, $SD = 10.36$; $p < .001$). The comparison with \emph{Combined Cues w/o Profile} did not survive correction ($M = 82.0\%$, $SD = 8.42$; $p = .021$).

The remaining contrasts clarify the role of each cue. \emph{Audio Only} outperformed \emph{Baseline} ($p < .001$), while \emph{Visual Only} did not outperform \emph{Baseline} after correction ($p = .030$). The two single-channel conditions did not differ ($p = .296$). \emph{Combined Cues w/o Profile} outperformed \emph{Baseline} and \emph{Visual Only}, but its comparison with \emph{Audio Only} also missed the corrected threshold ($p = .006$). Within this task, a name attached to the caption provided a more reliable attribution resource than a transient location marker alone.
\begin{figure}[t]
  \centering
  \includegraphics[width=\linewidth]{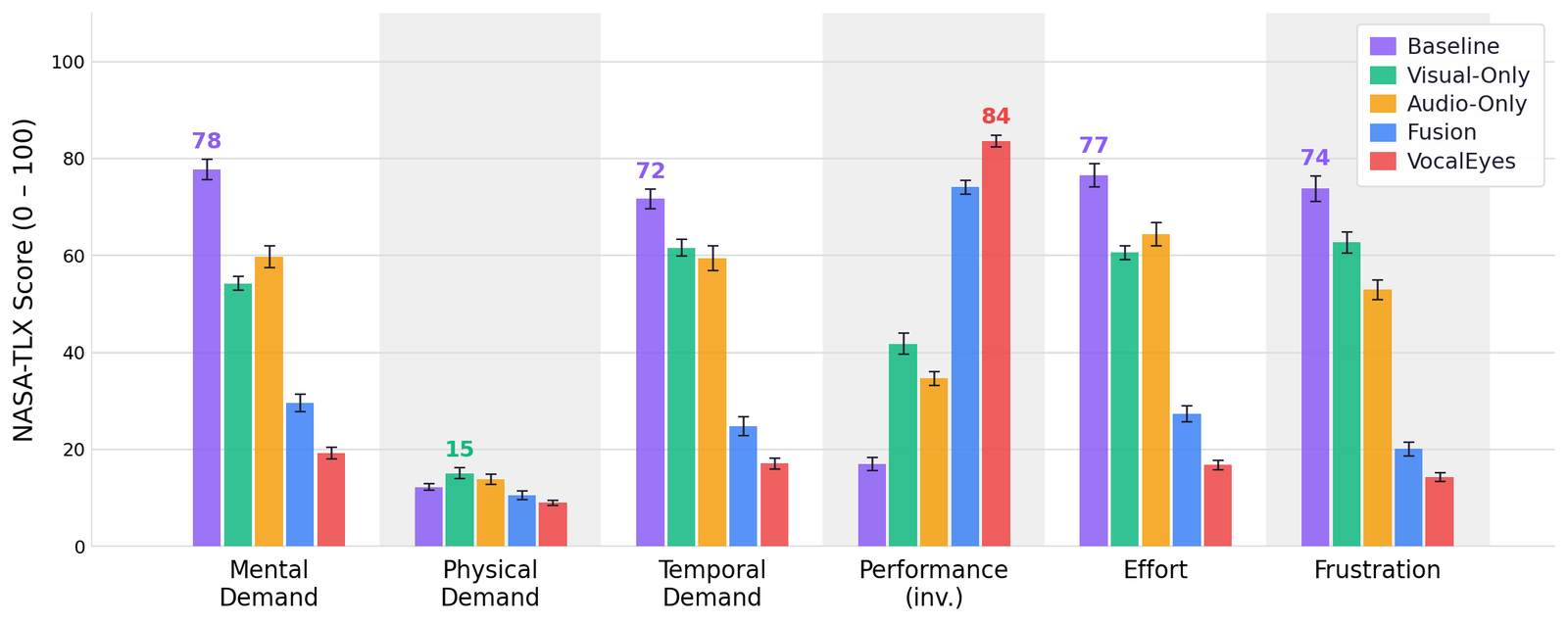}
  \vspace{-8mm}
  \caption{NASA-TLX scores across our study conditions. Lower scores indicate lower workload except for Performance, where higher scores indicate better performance.
  }
  \Description{Grouped bars compare six NASA-TLX dimensions across five conditions. VocalEyes has the lowest mental, physical, temporal, effort, and frustration scores and the highest performance score.}
  \label{fig:tlx}
  \vspace{-7mm}
\end{figure}

\subsubsection{The Complete Interface Reduced Workload (RQ3)}
Participants reported the lowest workload with the complete interface. \vocaleyes{} produced the lowest NASA-TLX score ($M = 32.6$, $SD = 5.01$; Figure~\ref{fig:tlx}) and outperformed \emph{Baseline} ($M = 72.3$, $SD = 9.20$; $p < .001$), \emph{Visual Only} ($M = 60.2$, $SD = 7.45$; $p < .001$), \emph{Audio Only} ($M = 54.2$, $SD = 8.74$; $p < .001$), and \emph{Combined Cues w/o Profile} ($M = 40.3$, $SD = 7.12$; $p = .007$). \emph{Visual Only} and \emph{Audio Only} did not differ after correction ($p = .094$). From \emph{Baseline} to \vocaleyes{}, Mental Demand fell from $77.8$ to $19.1$, Temporal Demand from $71.8$ to $17.1$, Effort from $76.5$ to $16.9$, and Frustration from $73.8$ to $14.4$ (all $p < .001$).

Participants connected this workload to the effort of reconstructing attribution in their written suggestions. P11 wrote about the caption-only condition: \textcolor{participant}{\textit{\textquotedblleft{}Without [the identification], I kept looking around trying to match voices to faces, and by the time I figured it out, I had missed what they actually said.\textquotedblright{}}} P14 contrasted that search with the complete interface: \textcolor{participant}{\textit{\textquotedblleft{}I could just focus on the content instead of playing detective.\textquotedblright{}}} These examples suggest that identity cues reduced a coordination task between reading and visual search. The comparison between \vocaleyes{} and \emph{Combined Cues w/o Profile} further indicates that the stable profile context affected experienced workload even though it did not change the recognition algorithm.

\subsubsection{Participants Valued the Complete Interaction and Meeting Record (RQ4)}
Usability ratings favored the complete interface. \vocaleyes{} produced the highest SUS score ($M = 85.9$, $SD = 5.52$; all pairwise $p \leq .002$), and all $20$ participants rated it above the threshold of $68$~\cite{brooke1996sus,bangor2008empirical}. The corresponding counts were $17$ for \emph{Combined Cues w/o Profile} ($M = 78.1$), $8$ for \emph{Audio Only} ($M = 64.5$), $5$ for \emph{Visual Only} ($M = 61.9$), and $1$ for \emph{Baseline} ($M = 55.0$). The two single-channel conditions did not differ ($p = 1.00$), and \emph{Visual Only} did not differ from \emph{Baseline}.

Feature ratings reveal both valued elements and remaining friction. Eighteen participants agreed or strongly agreed that they felt satisfied with \vocaleyes{}. All $20$ endorsed the meeting record, $17$ endorsed the caption bubble, and $16$ endorsed the directional indicator (Supplementary Figure~S5). Fourteen endorsed the identity label, while two disagreed. The downward arrow drew a more mixed response: $12$ participants found it non-distracting and six selected neutral. Three participants reported that caption latency disrupted their attention. Thus, participants valued the complete information structure while still noticing the cost of delayed captions and persistent visual elements.

Written suggestions also distinguished two post-meeting tasks. P8 compared the transcript's support for targeted retrieval with the avatar replay's support for reconstructing conversational sequence: \textcolor{participant}{\textit{\textquotedblleft{}[Avatar animation] is like watching a game replay---you see the whole picture. But if I just need a quote, I would go to the transcript.\textquotedblright{}}} P3 described how trust in live attribution changed during use: \textcolor{participant}{\textit{\textquotedblleft{}After the first couple of minutes, I stopped second-guessing it.\textquotedblright{}}} These examples show how participants connected immediate attribution and repeated system behavior with later inspection.

\section{Discussion}
Speaker-aware captioning depends on how an interface coordinates identity, location, and contextual metadata. We develop four study-grounded implications and one system-boundary implication. Each connects the evidence to an earlier design objective without repeating the condition means.

\noindent\textbf{\colorbox{fcolor}{F1}}~{\bf Preserve speaker attribution temporally.}
F1 connects RQ2 to DO1 and DO3. A spatial marker can orient attention toward the face that currently appears to articulate, but its meaning lasts only while that face remains visible and active. A name embedded in the caption persists with the utterance after the user looks elsewhere. The tracking pattern in \emph{RQ2} reflects this difference: identity-supported attribution was more consistent than location alone within the controlled task.

This distinction extends prior work on spatial speech bubbles and situated captions~\cite{peng2018speechbubbles,yamamoto2021see,zhang2022balloon}. Those interfaces strengthen the immediate spatial connection between text and a visible source. It also complements the findings from ChatMuse, where AR conversational support required users to divide attention between displayed information and multiple conversational partners, motivating support that adapts to the ongoing group interaction~\cite{zhou2026chatmuse}. Together, these findings highlight two complementary requirements: spatial cues can help users orient toward the current speaker, while persistent attribution preserves who produced an utterance after attention shifts elsewhere. \vocaleyes{} shows why a captioning system also needs a temporal connection. Users may read after the speaker finishes, revisit an earlier sentence, or inspect a record after the meeting. A view-dependent pointer cannot carry attribution across those transitions.

Designers should therefore bind the most durable identity cue to the most durable representation of the utterance. A speaker name belongs with the caption and transcript; a location cue belongs near the currently active face.  This extends ChatMuse's broader design consideration that AR support for group conversation should account for both verbal and nonverbal conversational context~\cite{zhou2026chatmuse}: rather than collapsing these signals into a single representation, speaker-aware interfaces can assign them complementary roles according to how long their information remains useful. The two cues can reinforce each other during a visible turn, but neither should stand in for the other. This principle also clarifies the modular design in DO3: when visual sensors lose the face, the interface can remove the marker without erasing a resolved name from the caption.

Large groups and overlapping speech require a different layout. Several simultaneous names can crowd a caption region, while a single pointer can oversimplify concurrent articulation. Per-speaker caption lanes, turn stacks, or explicit overlap representations can separate concurrent sources. This challenge becomes increasingly important as group size grows; as discussed in ChatMuse, the ``many minds'' problem makes listening, turn-taking, and interpreting conversational dynamics more difficult as additional participants join~\cite{zhou2026chatmuse}. Speaker--utterance relationships. Across these layouts, the interface should preserve each utterance's source beyond the instant when a detector fires.

\noindent\textbf{\colorbox{fcolor}{F2}}~{\bf Keep registration in the background and make identity changes visible.}
F2 connects DO2 to the controlled late-entry event in RQ1. Conventional speaker verification assumes that the system receives a named voice sample before use~\cite{dehak2010front,snyder2018x}. \vocaleyes{} instead accumulates acoustic evidence, waits for identity-bearing speech, and associates that evidence with a name. This approach is consistent with the broader direction explored in ChatMuse, where AR support is driven by conversational context that develops during the interaction rather than by a fixed representation established beforehand~\cite{zhou2026chatmuse}. Here, we extend this idea to speaker identity: information that emerges naturally during conversation can become part of the system state. The controlled event shows that a new participant can join without stopping the meeting for setup.

Before registration, \textsc{Unknown Speaker} describes the available evidence more accurately than a guessed identity. After registration, users should be able to inspect which utterance supplied the identity and correct the profile if semantic parsing or voice matching failed. This is especially important for conversational support because, as ChatMuse highlights, MR interfaces operate alongside an ongoing social interaction in which unnecessary or poorly timed information can itself become distracting~\cite{zhou2026chatmuse}. Registration feedback should therefore communicate meaningful state changes without requiring users to disengage from the conversation. The system should perform matching in the background, but the interface should show the resulting name and offer correction. This separates an unobtrusive process from a visible, repairable state change.

Designers can treat registration as a small state machine with visible transitions: \emph{unknown}, \emph{candidate identity}, \emph{registered}, and \emph{corrected}. Each transition should expose the evidence appropriate to the task. A live interface may show only a concise status, while a review interface can show the source utterance and subsequent matches. This separation between lightweight live feedback and richer retrospective information follows the same general concern raised in ChatMuse: conversational support should provide useful context while limiting the visual and attentional demands imposed during the live interaction~\cite{zhou2026chatmuse}. This structure supports recovery without forcing users to understand embeddings or similarity thresholds.

Natural introductions do not always contain complete or unambiguous identity information. People may introduce someone else, use a nickname, omit their organization, or rely on an existing social context. The system should therefore allow partial profiles and later enrichment rather than treating the first parse as final. It should also let users merge duplicate profiles, split a mistaken match, or return a profile to the unknown state. These repair actions turn unavoidable recognition errors into manageable interaction states.

\noindent\textbf{\colorbox{fcolor}{F3}}~{\bf Match each cue's duration to its purpose.}
Speaker-aware captioning combines outputs that arrive and expire at different rates. Visual mouth motion can update quickly, but it remains view-dependent and transient. Audio identity arrives after the system accumulates and processes an utterance, yet the resulting name can persist across captions and records. The profile card changes even more slowly because it supplies background context rather than moment-to-moment turn information. These differences suggest that conversational cues should not all be presented in the same way or for the same duration.

Multimodal fusion therefore involves an interface decision as well as a model-level decision. Audio-visual models can improve classification by combining evidence~\cite{tao2018gating,alcazar2020active,vasireddy2024robust}, but an interface still must decide how to represent the fused result. \vocaleyes{} keeps the channels separate and assigns each output a display role. The visual marker provides an immediate cue to the current speaker, the caption preserves speaker identity with the utterance, and the profile card provides longer-term context about an unfamiliar participant.

ChatMuse's study similarly showed that persistent AR support can require users to shift attention between displayed information and their conversational partners, motivating support that adapts to the ongoing conversation~\cite{zhou2026chatmuse}. The present results extend this consideration to speaker-aware captions and suggest a simple design principle: \emph{use transient cues for attention and persistent cues for context}. A spatial marker can direct attention to the current speaker and disappear when that evidence expires. Speaker identity can remain attached to the utterance, while profile information can persist when users still need context about an unfamiliar speaker.

The study evaluates the profile card as a bundle of name, role, and organization. A stable card can assist users in becoming familiar with an unfamiliar group, yet the same card can become redundant after repeated interaction or intrusive on a small display. Following ChatMuse's broader recommendation to reduce unnecessary visual support during conversation~\cite{zhou2026chatmuse}, future interfaces could collapse or hide profile information as speakers become familiar. Future studies should also isolate individual metadata fields to determine which information remains useful enough to display persistently.

\noindent\textbf{\colorbox{fcolor}{F4}}~
{\bf Keep meeting records traceable to the conversation.}
F4 connects the post-meeting responses in RQ4 to DO1. Participants used the transcript for targeted retrieval and the avatar replay to reconstruct conversational sequence. An incorrect live name can also become a lasting record of who made a statement. As speaker-aware information moves from live support to a persistent record, maintaining its link to the original discussion grows increasingly important.

This extends conversational support beyond the live interaction emphasized in systems such as ChatMuse~\cite{zhou2026chatmuse}. Information presented during a conversation can later support reviewing what occurred, who contributed an idea, or how the discussion developed. Designers should store attributions as inspectable provenance instead of immutable text. Each attribution record can link a person's name to the specific utterance, registration event, and the confidence level or status available at that time. This allows users to trace a stored attribution back to the interaction that produced it rather than treating the transcript as ground truth. Users should be able to correct individual utterances, update profiles, or propagate verified corrections across related exchanges. The interface must clearly differentiate between system-generated attributions and those confirmed by users, especially before summarizing or generating action items.

\noindent\textbf{\colorbox{fcolor}{F5}}~
{\bf Make speaker data boundaries clear to users.}
F5 follows from the system architecture and privacy literature rather than from the participant measures. The current prototype keeps raw audio and video within the local network but sends transcript text to a hosted language model for semantic parsing. It also retains speaker embeddings and attributed transcripts. Speaker-aware systems therefore involve several distinct data boundaries: what is captured, what is processed locally, what leaves the device or local network, and what remains after the meeting. A single \textquotedblleft{}on-device\textquotedblright{} label would collapse these distinct boundaries. Speaker-aware systems should instead explain where capture, inference, external transfer, and storage occur; they should request consent before retaining linkable voice profiles and provide deletion controls~\cite{rajaram2025privacy,nautsch2019preserving,tomashenko2022voiceprivacy}. A useful privacy design should therefore communicate not only \emph{what} data are collected, but also \emph{where} they go and \emph{how long} they persist.

Together, these findings describe how speaker identity should be managed throughout speaker-aware AR interaction. Speaker attribution should persist beyond transient spatial cues (\colorbox{fcolor}{F1}), while registration should occur naturally within conversation while remaining visible and correctable when needed (\colorbox{fcolor}{F2}). Information should remain available only while it is useful, with transient cues supporting attention and persistent cues providing identity and context (\colorbox{fcolor}{F3}). When conversational information becomes a meeting record, speaker attribution should remain traceable to the interaction that produced it (\colorbox{fcolor}{F4}). Finally, systems should make clear what speaker data are captured, where they are processed, and how long they are retained (\colorbox{fcolor}{F5}). Collectively, these findings position \vocaleyes{} as an exploration of how speaker identity can be registered, represented, maintained, and governed across the lifecycle of an AR-supported conversation.

\section{Limitations and Future Work}\label{sec:dcs}

Future work should extend the prototype along three directions: multimodal binding, latency, and interface evaluation.

The prototype has technical limits. Audio identity and face tracking remain separate, so overlapping speech can make their correspondence ambiguous. The amplitude-gated VAD reacts to loud non-speech sounds. MAR needs visible lip landmarks and can fail with pose, distance, occlusion, eating, or laughter. The English parser expects an explicit self-introduction. Future systems should bind utterances to face tracks over the same time interval, support overlap, abstain when evidence conflicts, and provide profile correction.

Latency and interface factors also need separate tests. The acoustic path takes about $1.5$--$4$\,s from speech onset to caption display. Figure~\ref{fig:latency} measures only the processing that starts after the segment closes. Streaming ASR could show a \emph{partial hypothesis}, or tentative text produced before an utterance ends, but the interface would need to revise unstable words~\cite{baumann2009incremental}. The comparison between \emph{Combined Cues w/o Profile} and \vocaleyes{} isolates the whole profile card, not its name, role, and organization fields. The post-meeting tools also received brief formative ratings rather than a task-based evaluation. Future work should test these choices independently and over repeated meetings.
\section{Conclusion}\label{sec:ccl}
\vocaleyes{} investigates how AR captions can preserve who said what when a meeting lacks pre-enrolled speaker profiles. The system turns natural self-introductions into named voice profiles and coordinates three user-visible layers: persistent attribution in captions, stable identity context, and transient spatial guidance. A controlled study with $20$ participants found that the complete interface supported more accurate speaker tracking and lower workload than caption-only AR in scripted small-group meetings. The component contrasts show that identity and location serve different interaction roles, while the post-meeting responses show why attribution must remain inspectable after the live conversation. In-conversation registration therefore offers a design strategy for acquiring identity during interaction and preserving its provenance afterward. Future speaker-aware interfaces should make registration visible, preserve uncertainty, support correction, and carry attribution provenance from live captions into meeting records.

\bibliographystyle{ACM-Reference-Format}
\bibliography{references}

\end{document}


\maketitle

This document contains additional system diagrams, parameter-selection evidence, and descriptive user-study results referenced by the main paper.

\begin{figure}[htbp]
    \centering
    \includegraphics[width=0.95\linewidth]{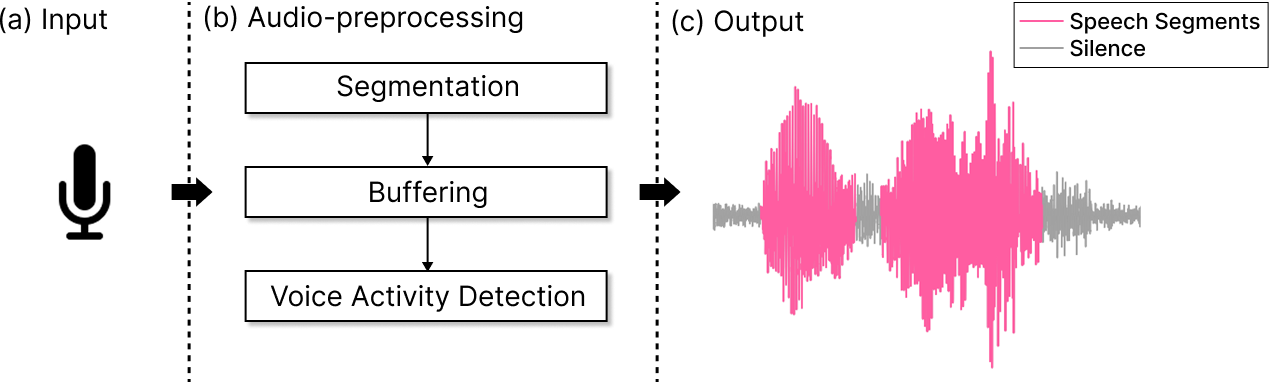}
    \caption{\vocaleyes{}'s audio pre-processing and speech-detection pipeline.}
    \Description{A block diagram shows segmentation, buffering, and voice activity detection beside an audio waveform. Pink marks speech, and gray marks silence.}
    \label{fig:supp_audio_preprocess}
\end{figure}

\begin{figure}[htbp]
    \centering
    \includegraphics[width=0.95\linewidth]{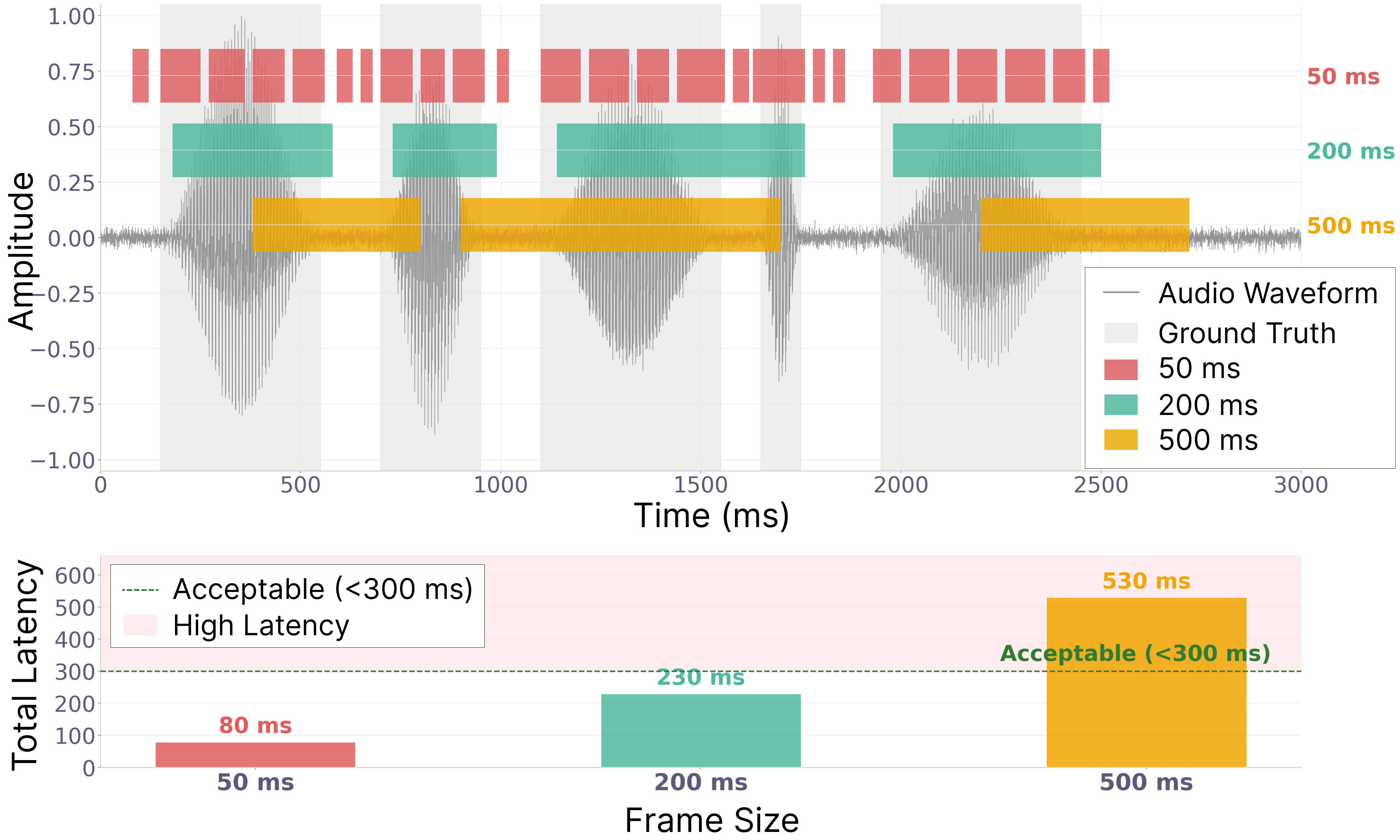}
    \caption{Frame-size selection analysis. (a) Colored bars compare detected speech with gray ground truth. Frames of $50$\,ms trigger on noise bursts; $500$\,ms frames merge turns and miss short segments; $200$\,ms provides the best pilot trade-off. (b) Bars show frame-trigger and segmentation overhead. The internal $300$\,ms target covers one $200$\,ms frame and one $100$\,ms client update; it does not represent a perceptual threshold or end-to-end caption latency.}
    \Description{Panel a overlays 50, 200, and 500 millisecond speech detections on a waveform and ground-truth speech regions. Panel b reports front-end overheads of 80, 230, and 530 milliseconds, with the 200 millisecond setting below the 300 millisecond engineering target.}
    \label{fig:supp_frame_analysis}
\end{figure}

\begin{figure}[htbp]
    \centering
    \includegraphics[width=0.42\linewidth]{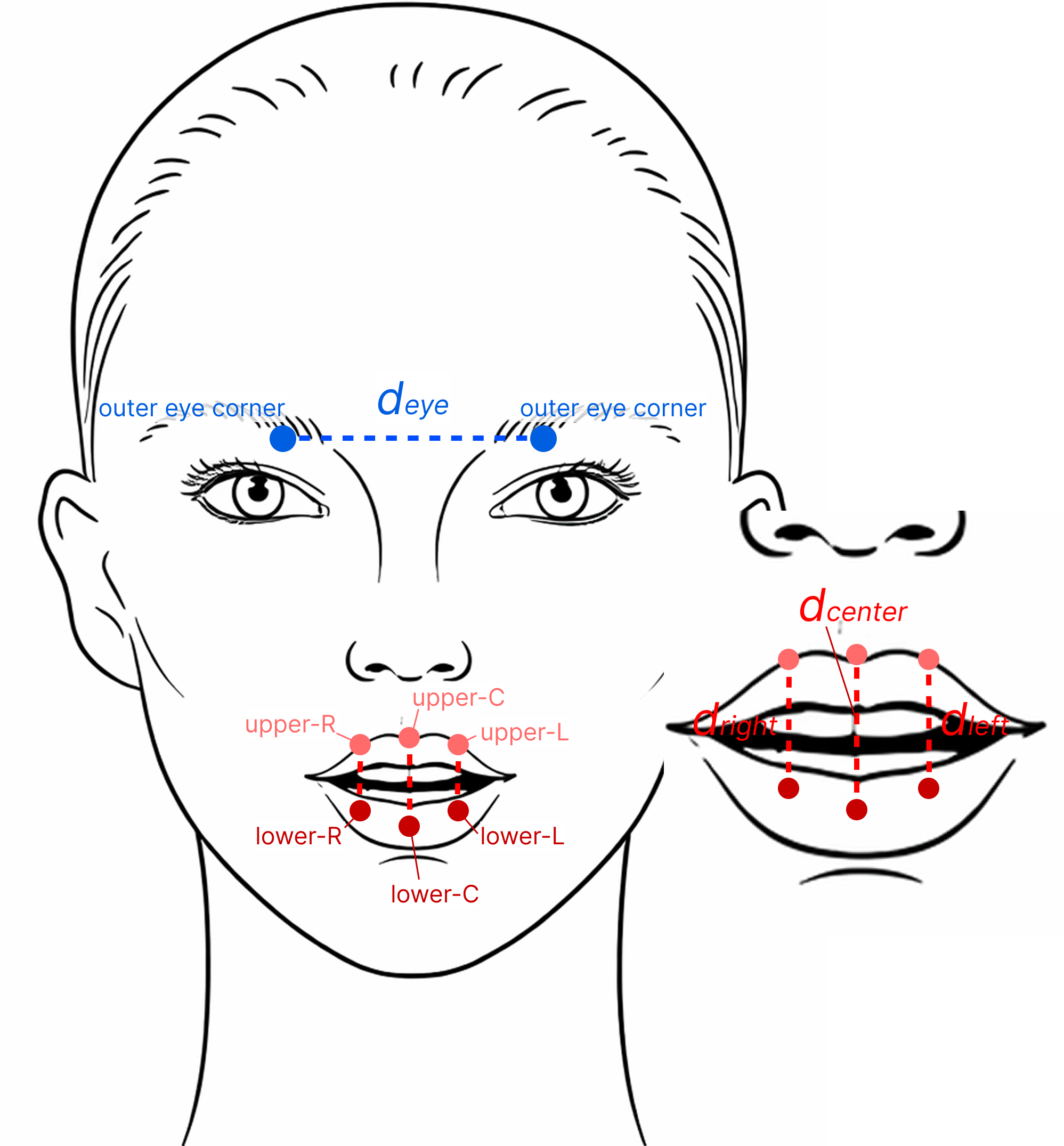}
    \caption{Mouth aspect ratio (MAR) computation: $\text{MAR} = \frac{d_\text{center} + d_\text{left} + d_\text{right}}{3 \cdot d_{\text{mouth}}}$.}
    \Description{A close-up diagram of a mouth marks three vertical distances between paired upper and lower lip landmarks and one horizontal distance between the mouth corners.}
    \label{fig:supp_mar}
\end{figure}

\begin{figure}[htbp]
    \centering
    \includegraphics[width=0.95\linewidth]{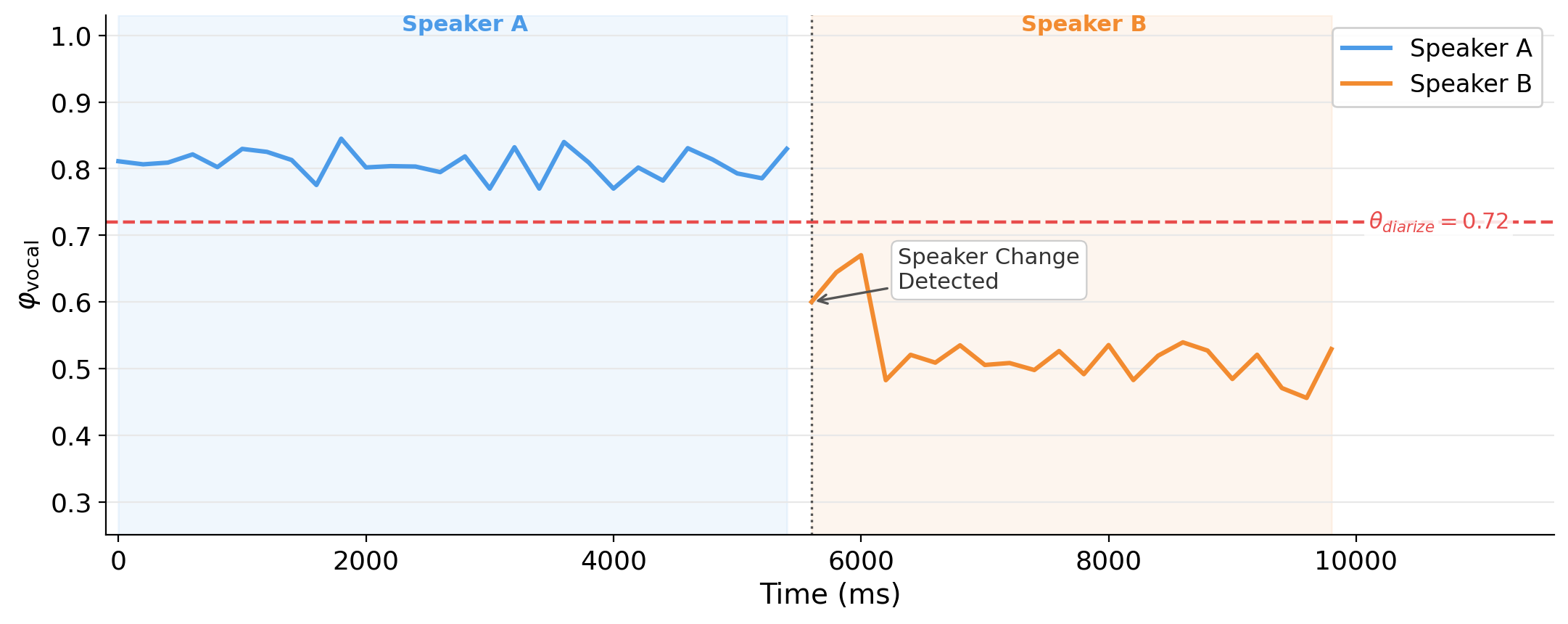}
    \caption{$\varphi_\text{vocal}$ during a representative two-speaker session. Similarity against $\mathbf{e}_k$ stays high during a stable turn and drops below $\theta_\text{vocal} = 0.72$ at the transition. The system then re-anchors $\mathbf{e}_k$ to the incoming speaker.}
    \Description{A time-series plot shows vocal similarity above the $0.72$ threshold during one speaker's turn and a drop at the speaker change. The curve rises after the system re-anchors the reference embedding.}
    \label{fig:supp_speaker_change}
\end{figure}

\begin{figure}[htbp]
    \centering
    \includegraphics[width=0.95\linewidth]{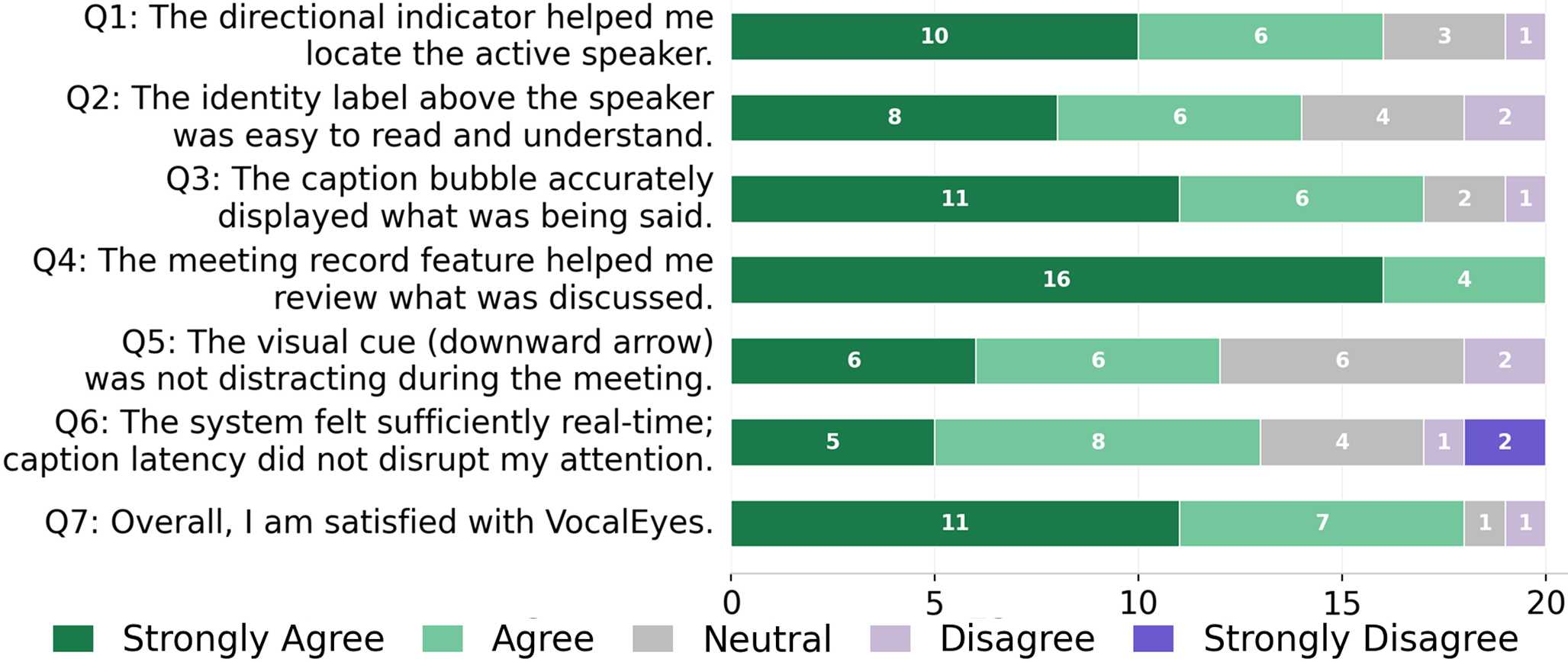}
    \caption{Feature satisfaction ratings. The stacked bars show responses across seven system features ($N = 20$); each segment gives the number of participants who selected that rating.}
    \Description{A horizontal stacked bar chart reports five-point satisfaction responses from 20 participants for seven VocalEyes features, with most responses concentrated in Agree and Strongly Agree.}
    \label{fig:supp_features}
\end{figure}

\section{Rule-Based Parsing Pattern}

Table~\ref{tab:supp_pattern} lists the spaCy NER and regular-expression rules that serve as a fallback when LLM extraction fails.

\begin{table}[htbp]
  \centering
  \setlength{\tabcolsep}{4pt}
  \footnotesize
  \begin{tabularx}{\linewidth}{@{}p{10mm} p{22mm} X p{34mm}@{}}
    \toprule
    \textbf{Field} & \textbf{Method} & \textbf{Pattern or trigger} & \textbf{Example} \\
    \midrule
    \multirow{2}{*}{Name}
      & spaCy NER & PERSON entity
      & \textit{Dr.\ Sarah Johnson} \\[2pt]
      & Regex & \texttt{(?:I'm|I~am|my~name~is)} \newline
               \texttt{\textbackslash s+([A-Z][a-zA-Z'-]+} \newline
               \texttt{(?:\textbackslash s+[A-Z][a-zA-Z'-]+)\{0,2\})}
      & \textit{My name is Lily Chen} \\
    \addlinespace[4pt]
    \multirow{2}{*}{Org}
      & spaCy NER & ORG entity
      & \textit{Stanford Univ.} \\[2pt]
      & Regex & \texttt{(?:at|in|from)\textbackslash s+} \newline
               \texttt{([A-Z][a-zA-Z\&'-]+} \newline
               \texttt{(?:\textbackslash s+[A-Z][a-zA-Z\&'-]+)*)}
      & \textit{I work at Bright} \newline \textit{Future Tech} \\
    \addlinespace[4pt]
    \multirow{2}{*}{Role}
      & Dependency & Attributive token with ``be'' head
      & \textit{I am a researcher} \\[2pt]
      & Regex & \texttt{(?:a|an)\textbackslash s+} \newline
               \texttt{((?:[A-Za-z-/]+\textbackslash s+)\{0,2\}} \newline
               \texttt{[A-Za-z-/]+?)} \newline
               \texttt{\textbackslash s+(?:at|from|in)}
      & \textit{a senior engineer at X} \\
    \bottomrule
  \end{tabularx}
  \caption{Rule-based fallback parsing. The parser first applies spaCy NER and then applies a regular-expression fallback for each field.}
  \label{tab:supp_pattern}
\end{table}

\section{Meeting Timeline}

\begin{figure}[htbp]
  \centering
  \includegraphics[width=\linewidth]{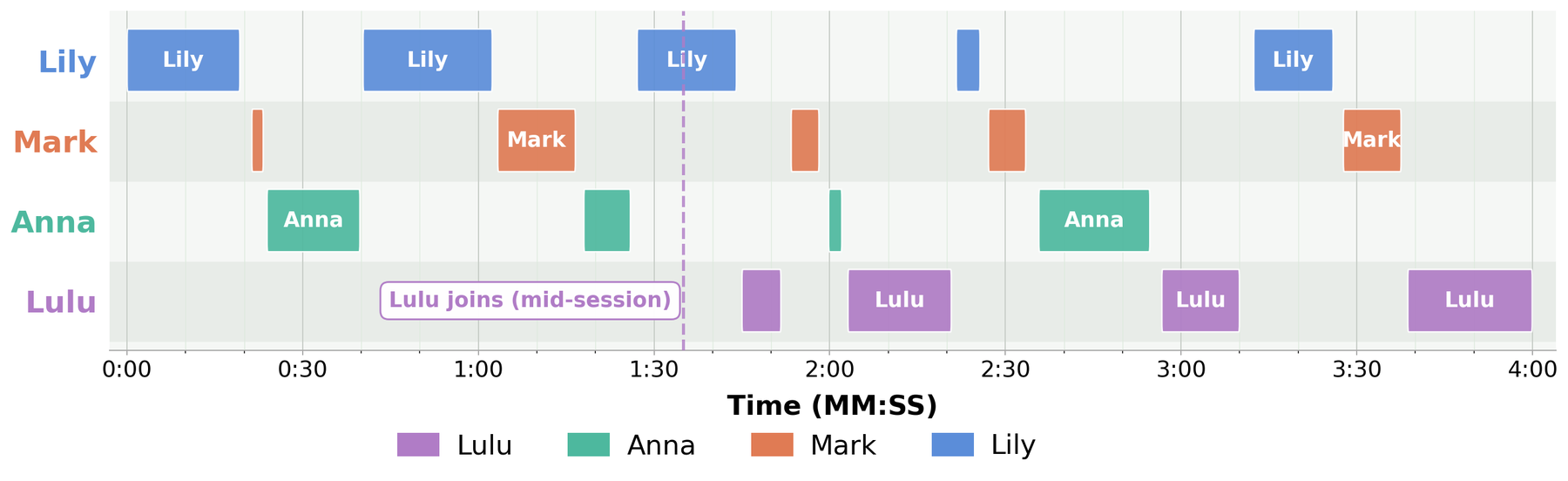}
  \caption{Representative meeting timeline in the \vocaleyes{} condition. Each row represents one speaker, and labeled color blocks show system detections. Lulu joined midway, and the system registered her without pausing the conversation.}
  \Description{A multi-row timeline plots color-coded speaking intervals for the participant and confederates. Lulu's row begins around 1 minute 35 seconds and continues with later detected turns.}
  \label{fig:supp_timeline}
\end{figure}

Figure~\ref{fig:supp_timeline} shows one \vocaleyes{} session. Lulu joined and introduced herself at about $1{:}35$, which triggered in-conversation registration.

\section{Vocal Similarity Distributions}

\begin{figure}[htbp]
  \centering
  \includegraphics[width=\linewidth]{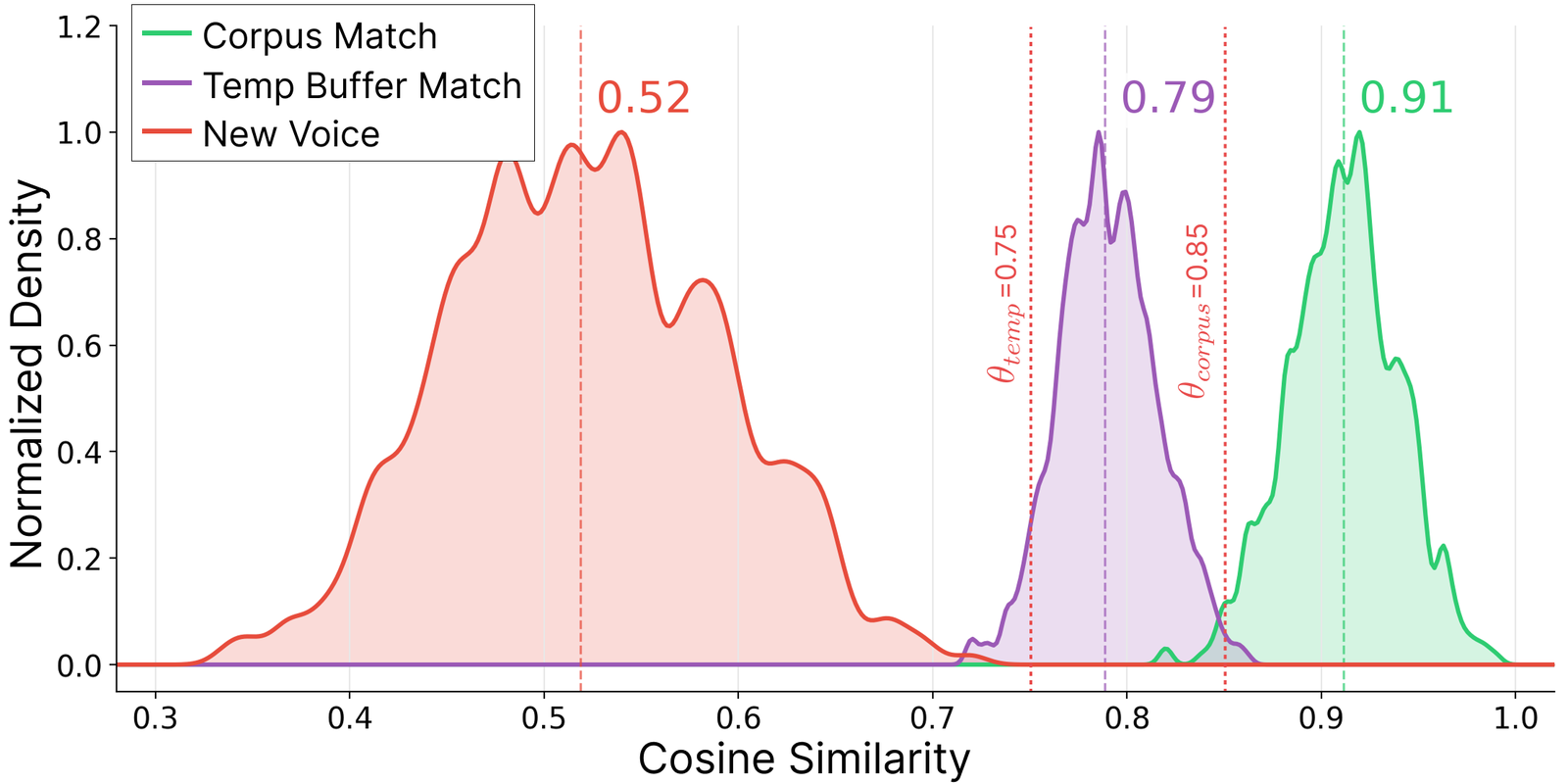}
  \caption{Vocal similarity $\varphi_{\text{vocal}}$ in the pilot recordings. Registered-profile matches occupy the highest range, temporary-buffer matches occupy the middle range, and new voices occupy the lowest range. Their separation guided the $0.75$ and $0.85$ thresholds.}
  \Description{Three distributions compare vocal-similarity values for a new voice, a temporary-buffer match, and a registered-profile match. Vertical lines mark thresholds at 0.75 and 0.85.}
  \label{fig:supp_similarity}
\end{figure}

Figure~\ref{fig:supp_similarity} shows the three pilot similarity distributions. Their separation guided $\theta_\text{temp} = 0.75$ and $\theta_\text{corpus} = 0.85$.

\section{Participant Demographics}

The sample included $8$ undergraduates, $7$ master's students, and $5$ doctoral students from Computer Science, Information Science, Psychology, and Design. Eleven participants reported normal vision, six wore glasses, and three wore contact lenses. Six participants who usually wore glasses used contact lenses during the study. Prior AR or VR experience ranged from zero to four years ($M = 1.2$, $SD = 1.1$). No participant had used Magic Leap~2. Eight participants spoke English natively, $10$ reported fluent non-native proficiency, and two reported intermediate proficiency.

\section{Feature Satisfaction}

Figure~\ref{fig:supp_features} reports seven feature ratings on a five-point scale. Eighteen participants agreed or strongly agreed that they felt satisfied with the system. All $20$ endorsed the meeting record, while $17$ endorsed the caption bubble and $16$ endorsed the directional indicator. Fourteen endorsed the identity label, and two disagreed. Twelve participants found the downward arrow non-distracting, and six selected neutral. Three participants reported that caption latency disrupted their attention.